\documentclass[twocolumn]{aastex63}

\received{October 23, 2020}
\revised{March 23, 2021}
\accepted{April 11, 2021}
\submitjournal{the Astronomical Journal}

\shorttitle{The Age--Metallicity--Specific Orbital Energy Relation for
MW Globular Clusters}

\shortauthors{Woody \& Schlaufman}

\begin{document}

\title{The Age--Metallicity--Specific Orbital Energy Relation for the
Milky Way's Globular Cluster System Confirms the Importance of Accretion
for Its Formation}

\correspondingauthor{Turner Woody}
\email{twoody1@jhu.edu}

\author[0000-0002-0721-6715]{Turner Woody}
\affiliation{Department of Physics and Astronomy \\
Johns Hopkins University \\
3400 N Charles St \\
Baltimore, MD 21218, USA}

\author[0000-0001-5761-6779]{Kevin C. Schlaufman}
\affiliation{Department of Physics and Astronomy \\
Johns Hopkins University \\
3400 N Charles St \\
Baltimore, MD 21218, USA}

\begin{abstract}

\noindent
Globular clusters can form inside their host galaxies at high redshift
when gas densities were higher and gas-rich mergers were common.  They can
also form inside lower-mass galaxies that have since been accreted and
tidally disrupted, leaving their globular cluster complement bound to
higher-mass halos.  We argue that the age--metallicity--specific orbital
energy relation in a galaxy's globular cluster system can be used to
identify its origin.  Gas-rich mergers should produce tightly bound
systems in which metal-rich clusters are younger than metal-poor clusters.
Globular clusters formed in massive disks and then scattered into a halo
should have no relationship between age and specific orbital energy.
Accreted globular clusters should produce weakly bound systems in
which age and metallicity are correlated with each other but inversely
correlated with specific orbital energy.  We use precise relative ages,
self-consistent metallicities, and space-based proper motion-informed
orbits to show that the Milky Way's metal-poor globular cluster system
lies in a plane in age--metallicity--specific orbital energy space.
We find that relatively young or metal-poor globular clusters are weakly
bound to the Milky Way, while relatively old or metal-rich globular
clusters are tightly bound to the Galaxy.  While metal-rich globular
clusters may be formed either in situ or ex situ, our results suggest that
metal-poor clusters formed outside of the Milky Way in now-disrupted dwarf
galaxies.  We predict that this relationship between age, metallicity,
and specific orbital energy in a $L^{*}$ galaxy's globular cluster system
is a natural outcome of galaxy formation in a $\Lambda$CDM universe.

\end{abstract}


\section{Introduction}\label{intro}

Globular cluster formation is one the enduring unsolved problems
in astrophysics.  Globular clusters can form either in situ
inside their parent galaxy or ex situ in lower-mass galaxies that
have been accreted and tidally disrupted by a more massive halo
\citep[e.g.,][]{West_2004,Forbes_2018}.  In either case, the ancient
ages inferred for most globular clusters indicate that they formed long
ago when the universe was very different from its current $z = 0$ state.
Indeed, the interval spanning the 16th and 84th percentiles of the Milky
Way globular cluster system's age distribution is $11~\text{Gyr} \lesssim
\tau \lesssim 13~\text{Gyr}$ or $2.4 \lesssim z \lesssim 7.3$ in terms
of redshift \citep[e.g.,][]{Marin_2009,VandenBerg_2013,Wright_2006}.
Both in situ and ex situ formation channels are theoretically
expected to contribute to the Milky Way's globular cluster system
\citep[e.g.,][]{Griffen_2010,Renaud_2017}.

In the in situ formation scenario, globular clusters
form inside their parent galaxy in gas-rich galaxy mergers
\citep[e.g.,][]{Ashman_1992,Muratov_2010,Li_2014,Li_2019,Kim_2018}
and/or in gas-rich disks
\citep[e.g.,][]{Kravtsov_2005,Kruijssen_2015,Pfeffer_2018,Keller_2020}.
Galaxies in general and galaxies destined to be similar to the Milky Way
in particular are observed to be smaller at high redshift when globular
clusters form in this scenario \citep[e.g.,][]{Patel_2013a,Patel_2013b}.
Consequently, globular clusters formed in this way should be tightly
bound to their now more-massive and larger parent galaxies at $z = 0$
\citep[e.g.,][]{Leaman_2013}.  Another robust prediction of the gas-rich
merger scenario is that metal-rich clusters should be systematically
younger than metal-poor clusters \citep{Li_2014,Li_2019}.

In the ex situ scenario, globular clusters form
outside their $z = 0$ host halo in parent galaxies
that have since been accreted and tidally disrupted
\citep[e.g.,][]{Searle_1978,Mackey_2004,Forbes_2010,Forbes_2020,Massari_2019}.
While in most cases it is difficult to conclusively associate a candidate
accreted globular cluster with its parent galaxy, the globular clusters
Arp 2, NGC 6715 (M 54), Pal 12, Terzan 7, and Terzan 8 have all been
securely associated with the Sagittarius dwarf spheroidal (dSph) galaxy
\citep[e.g.,][]{Law_2010,Sohn_2018}.  Given the collisionless nature of
their accretion, globular clusters formed ex situ should be on average
less tightly bound to their host galaxy at $z = 0$ than clusters formed
in situ in gas-rich conditions.

The observed properties of accreted globular clusters can be used to
set limits on the properties of their parent galaxies.  Likewise, the
ensemble properties of a galaxy's accreted globular cluster system
can be used to explore its formation.  The maximum metallicity
realized in a dwarf galaxy's globular cluster system should not
exceed the dwarf galaxy's typical metallicity.  This is the case in
both the Fornax and Sagittarius dSph galaxies.  The Fornax dSph has
$\langle[\text{Fe/H}]\rangle = -1.04$ \citep{Kirby_2013}, while its
five globular clusters span the range $-2.5 \lesssim [\text{Fe/H}]
\lesssim -1.4$ \citep[e.g.,][]{Letarte_2006,Larsen_2012}.
While the Sagittarius dSph has $\langle[\text{Fe/H}]\rangle
\approx -0.5$ \citep[e.g.,][]{Chou_2007,Hasselquist_2017},
the five globular clusters securely associated with it span
the metallicity range $-2.3 \lesssim [\text{Fe/H}] \lesssim -0.6$
\citep[e.g.,][]{Cohen_2004,Sbordone_2007,Mottini_2008,Carretta_2010,Carretta_2014}.
A lower limit on the stellar mass of a galaxy necessary for the formation
of a globular cluster of a given metallicity can therefore be derived
from a stellar mass--metallicity relationship.  \citet{Kirby_2013} showed
that there exists a linear relationship between log stellar mass and
metallicity for dwarf galaxies over five orders of magnitude such that
the typical metallicity of a dwarf galaxy rises linearly with log stellar
mass from $[\text{Fe/H}] \approx -2.5$ at $M_{\ast} \sim 10^{3}~M_{\odot}$
to $[\text{Fe/H}] \approx -1.0$ at $M_{\ast} \sim 10^{8}~M_{\odot}$.  So
while both metal-rich and metal-poor globular clusters can form in massive
dSph galaxies, low-mass dSph galaxies will only be able to form metal-poor
clusters.  More sophisticated arguments based on cosmological dark
matter-only or hydrodynamical simulations have reached similar conclusions
\citep[e.g.,][]{Choksi_2018,Kruijssen_2019a,Kruijssen_2019b,Kruijssen_2019c}.

The mass of an accreted globular cluster's parent galaxy will also
affect its orbit.  A dwarf galaxy orbiting inside the dark matter halo
of a more massive galaxy will lose angular momentum due to dynamical
friction and its orbit will decay on a timescale that is proportional
to its mass \citep[e.g.,][]{Binney_2008}.  Massive dSph galaxies
will therefore be dragged into the inner regions of their host halo
and tidally disrupted much more quickly than low-mass dSph galaxies.
In particular, a massive dSph galaxy resembling Fornax with a total
mass $M_{\mathrm{tot}} \sim 10^{8}~M_{\odot}$ capable of forming a
globular cluster with $[\mathrm{Fe/H}] \approx -1.0$ will be tidally
disrupted 100 times faster than a low-mass dSph galaxy with a total
mass $M_{\mathrm{tot}} \sim 10^{6}~M_{\odot}$ that could not form a
globular cluster more metal-rich than $[\mathrm{Fe/H}] \approx -2.0$
\citep[e.g.,][]{Pascale_2018,Simon_2019}.  The metal-rich globular
clusters deposited deep inside the more massive halo when their parent
massive dSph galaxy is tidally disrupted will be much more tightly bound
to the massive halo than any metal-poor globular clusters left behind
when their parent low-mass dSph galaxy is tidally disrupted much later.

The ages of accreted globular clusters should be related to the masses
and therefore the metallicities of their parent galaxies too.  In similar
environments, galaxies destined to be massive form in higher $\sigma$
peaks in the primordial matter density distribution than galaxies destined
to be low mass \citep[e.g.,][]{Mo_2010}.  Since the dynamical time
$t_{\mathrm{dyn}}$ scales like $t_{\mathrm{dyn}} \propto \rho^{-1/2}$,
in similar environments more massive galaxies form stars earlier and
therefore the ages of the oldest stellar populations in galaxies should
be correlated with their masses.  The implication is that in similar
environments the age of the oldest globular cluster formed in a dwarf
galaxy should be correlated with its mass.  This is the case for the
$M_{\mathrm{tot}} \approx 9 \times 10^{7}~M_{\odot}$ Fornax dSph galaxy,
the $M_{\mathrm{tot}} \gtrsim 4 \times 10^{8}~M_{\odot}$ Sagittarius
dSph galaxy, and the $M_{\mathrm{tot}} \sim 10^{11}~M_{\odot}$ Large
Magellanic Cloud \citep[e.g.,][]{Pascale_2018,Vasiliev_2020,Erkal_2019}.
The Fornax globular clusters 1, 3, and 5 are comparable
in age to NGC 4590 (M 68) while cluster 4 is somewhat younger
\citep[e.g.,][]{Buonanno_1998,Buonanno_1999}.  On the other hand, Terzan 8
is the oldest globular cluster with a confirmed Sagittarius association
and is older than NGC 4590 \citep[e.g.,][]{Marin_2009,Sohn_2018}.
Likewise, the oldest of the Large Magellanic Cloud's globular
clusters that are not affected by crowding, Hodge 11, is older than
Terzan 8 but younger than the Milky Way's oldest globular clusters
\citep{Wagner-Kaiser_2017}.  Analyses based on scaling relations
calibrated with cosmological hydrodynamical simulations reach similar
conclusions \citep[e.g.,][]{Kruijssen_2019a}.

The relationship between age and metallicity in the Galaxy's globular
cluster system has long been used to constrain Milky Way formation.
As we described above, analyses of the relationship between age,
metallicity, and orbital properties in the Milky Way's globular cluster
system can be even more informative.  If globular clusters formed in
situ are produced in gas-rich mergers, then metal-rich clusters should
be younger than metal-poor clusters and both metal-rich and metal-poor
clusters should be relatively tightly bound to their parent galaxy.
If globular clusters are formed in situ in gas-rich disks and subsequently
scattered into the halo, then there should be no relationship between
age and specific orbital energy.  In the ex situ formation and accretion
scenario, older, metal-rich globular clusters should be tightly bound to
their $z = 0$ host galaxies while younger, metal-poor globular clusters
should be weakly bound to their $z = 0$ host.  These straightforward
arguments have been confirmed by cosmological hydrodynamical simulations
\citep[e.g.,][]{Pfeffer_2020}.

In this paper, we calculate the orbits of the Milky Way's globular
clusters with precise space-based astrometry.  We use those orbits
along with precise relative ages and metallicities from the literature
to quantify the relationship between age, metallicity, and specific
orbital energy in the Milky Way's globular cluster system.  We describe
the construction of our globular cluster sample in Section \ref{data}.
We detail in Section \ref{analysis} our orbital integrations and
statistical analyses of the age--metallicity--specific orbital energy
relation in our complete sample of globular clusters as well as its
metal-rich/metal-poor subsamples.  We review our results and their
implications for the formation of the Milky Way's globular cluster
system specifically and globular cluster formation in general in Section
\ref{discussion}.  We conclude by summarizing our findings in Section
\ref{conclusion}.

\section{Data}\label{data}

In an analysis of the age--metallicity--specific orbital energy
relation in the Milky Way's globular cluster system, the limiting
factors are precise cluster ages and proper motions.  Most of
our proper motion measurements come from Gaia Data Release 2 \citep[DR2;
][]{Gaia_2016,Gaia_2018a,Gaia_2018b,Crowley_2016,Fabricius_2016,Arenou_2018,Hambly_2018,Lindegren_2018,Luri_2018}.
We supplement the globular cluster proper motions published in
\citet{Gaia_2018b} with the Gaia DR2-based proper motions for NGC 6584 and
NGC 6723 from \citet{Baumgardt_2019} that have proper motion precisions
$\mu/\sigma_{\mu} > 85$.\footnote{We use this threshold because all
of the globular cluster proper motions from \citet{Gaia_2018b} have
$\mu/\sigma_{\mu} > 85$ in the \citet{Baumgardt_2019} catalog.} We also
add proper motions measurements for NGC 1261, NGC 4147, NGC 6101, Terzan
7, Arp 2, Terzan 8, NGC 6934, and Pal 12 from the High-resolution Space
Telescope Proper Motion Collaboration \citep[HSTPROMO; ][]{Sohn_2018}.  We
list our input globular cluster proper motions in Table \ref{tab:input}.

Globular cluster age determinations are inherently difficult to put
on an absolute age scale and can be non-trivially affected by even
small differences in analysis techniques.  Since the analyses we
will describe in Section \ref{analysis} only make use of relative
age differences, we choose to use the homogeneous relative globular
cluster ages presented in \citet{Marin_2009} to avoid the difficulties
associated with absolute ages.  \citet{Marin_2009} calculated
relative clusters ages $\tau = \text{age}/\text{12.8~Gyr}$ for two
different metallicity scales \citep{Zinn_1984,Carretta_1997}
as reported in \citet{Rutledge_1997b,Rutledge_1997a}
and four different sets of theoretical stellar models
\citep{Bertelli_1994,Girardi_2000,Pietrinferni_2004,Dotter_2007}.
We list our adopted globular cluster metallicities and normalized ages
in Table \ref{tab:inference}.  While we focus on the relative ages
calculated assuming the more modern \citet{Carretta_1997} metallicity
scale and \citet{Dotter_2007} models given in Table \ref{tab:inference},
the analysis we describe in Section \ref{analysis} reaches the same
conclusion regardless of the assumed metallicity scale or model grid.

We adopt the globular cluster distances and radial
velocities listed in Table \ref{tab:input} from
the December 2010 revision of the \citet{Harris_2010}
compilation.\footnote{\url{https://physwww.mcmaster.ca/~harris/mwgc.ref}}
Since the December 2010 revision of the \citet{Harris_2010} compilation
does not provide uncertainties for its distance estimates, we estimate
distance uncertainties by first transforming the distance inferences into
distance moduli.  We next assume a uniform distance modulus uncertainty
of 0.05 mag and sample each cluster's distance modulus from a normal
distribution centered on its distance modulus with standard deviation
0.05 mag.  We then transform the simulated distance moduli calculated
in this way back into distances and uncertainties.  We will use the
57 Milky Way globular clusters listed in Tables \ref{tab:input} and
\ref{tab:inference} with homogeneous relative ages and metallicities plus
precise space-based proper motions to explore the relationship between
age, metallicity, and specific orbital energy in Section \ref{analysis}.

\section{Analysis}\label{analysis}

We use the right ascensions $\alpha$, declinations $\delta$, proper
motions $\mu_{\alpha}$ \& $\mu_{\delta}$, distances $d$, and radial
velocities $v_{r}$ given for the 57 globular clusters listed in Table
\ref{tab:input} to calculate their orbits and specific orbital energies.
We use a Monte Carlo approach to average over the uncertainties in
our input proper motions, distances, and radial velocities.  On each
iteration, we sample the necessary input data for an orbit integration
($\alpha$, $\delta$, $\mu_{\alpha}$, $\mu_{\delta}$, $d$, and $v_{r}$)
from the data and uncertainties in Table \ref{tab:input}.  When possible,
we account for the covariance between simultaneously inferred proper
motion and parallax inferences.  For the globular clusters with proper
motions from \citet{Gaia_2018b}, we account for the covariance between
proper motions and parallaxes even though we use \citet{Harris_2010}-based
distances instead of the Gaia DR2 parallax-based distances.  For the
globular clusters with proper motions from \citet{Baumgardt_2019},
we account for the covariance between the proper motion components.
\citet{Sohn_2018} did not provide estimates of the covariance between
proper motion components, so we sample each proper motion component
independently from within its quoted uncertainty.  For distances and
radial velocities, we sample from normal distributions with the means
and standard deviations given in Table \ref{tab:input}.  We repeat this
process 100 times to generate 100 sets of initial conditions for each
globular cluster.

For each Monte Carlo initial condition realization, we integrate a
cluster's orbit in model Milky Way potentials using the \texttt{galpy}
python module\footnote{\url{https://github.com/jobovy/galpy}}
\citep{Bovy_2015}.  We calculate orbits from the present five Gyr
into the past with a time step of 200,000 years assuming a solar
motion with respect to the local standard of rest ($U_{\odot}$,
$V_{\odot}$, $W_{\odot}$) = (11.1, 12.24, 7.25) km s$^{-1}$
\citep[e.g.,][]{Schonrich_2010}.

We calculate orbits in three different model Milky Way potentials:
the default \texttt{MWPotential2014} from \citet{Bovy_2015}, a scaled
version of \texttt{MWPotential2014} with its halo mass increased by 50\%,
and \texttt{McMillan17} from \citet{McMillan_2017}.  In the default
\texttt{MWPotential2014} potential, the virial mass and radius are
$M_{\text{vir}} = 0.8 \times 10^{12}~M_{\odot}$ and $r_{\text{vir}}
= 245$ kpc respectively.  Its bulge is parameterized as a power-law
density profile with exponent $-1.8$ that is exponentially cut-off at
1.9 kpc.  Its disk is represented by a Miyamoto--Nagai potential with
a radial scale length of 3 kpc and a vertical scale height of 280 pc
\citep{Miyamoto_1975}.  Its halo is modeled as a Navarro--Frenk--White
(NFW) halo with a scale length of 16 kpc \citep{Navarro_1996}.  The scaled
version of the \texttt{MWPotential2014} potential differs from the default
\texttt{MWPotential2014} potential in that we have increased its halo
mass by 50\%.  This change increases the virial mass to $M_{\text{vir}} =
1.2 \times 10^{12}~M_{\odot}$.  In both cases, we set the solar distance
to the Galactic center to $R_{0}=8.122$ kpc, the circular velocity at
the Sun to $V_{0}=238$ km s$^{-1}$, and the height of the Sun above the
plane to $z_{0}=25$ pc \citep{Juric_2008,Bland_2016,Gravity_2018}.

The \texttt{McMillan17} potential has $M_{\text{vir}} = 1.4 \times
10^{12}~M_{\odot}$.  Its bulge is parameterized as a spherical power
law density profile with exponent $-1.8$ that is finite at $r = 0$ and
has an exponential cut-off at 2.1 kpc.  Its disk is split into four
components: a thin stellar disk, a thick stellar disk, an \ion{H}{1}
gas disk, and an H$_{2}$ gas disk.  Its two stellar disks are modeled
as exponential in both the cylindrical radius $R$ and height $z$, with
scale lengths $R_{\text{d,thin}}=2.6$ kpc, $R_{\text{d,thick}}=3.6$ kpc,
$z_{\text{d,thin}}=300$ pc, and $z_{\text{d,thick}}=900$ pc.  Its two
gas disks are modeled as exponential in $R$ with a central hole and a
modified exponential profile in the $z$ coordinate.  Its halo is modeled
as a NFW halo with a scale length of 19.6 kpc.  The \texttt{McMillan17}
potential uses a solar distance and circular velocity of 8.21 kpc and
233.1 km s$^{-1}$.

The Milky Way's mass within 25 kpc $M_{25}$ is especially relevant
for this problem, and estimates of $M_{25}$ for the Milky Way
based on globular cluster kinematics suggest that $M_{25} \approx
0.26_{-0.06}^{+0.10} \times 10^{12}~M_{\odot}$ \citep{Eadie_2019}.
The $M_{25}$ values for the default \texttt{MWPotential2014}, the
scaled \texttt{MWPotential2014}, and the \texttt{McMillan17} potentials
are $M_{25} = 0.25 \times 10^{12}~M_{\odot}$, $M_{25} = 0.29 \times
10^{12}~M_{\odot}$, and $M_{25} = 0.28 \times 10^{12}~M_{\odot}$
respectively.  All three potentials are therefore consistent the
best available Milky Way constraints in the same volume.  While
both the default \texttt{MWPotential2014} and \texttt{McMillan17}
potentials are generally consistent with literature estimates of
the Milky Way's virial mass, most recent mass estimates making use
of Gaia DR2 proper motions suggest virial masses in the range $1.1
\times 10^{12}~M_{\odot} \lesssim M_{\text{vir}} \lesssim 1.4 \times
10^{12}~M_{\odot}$ \citep[e.g.,][]{Wang_2020}.  While the virial mass
of the \texttt{McMillan17} potential is in this interval, the virial
mass of the default \texttt{MWPotential2014} potential is below this
range.  In addition, the \texttt{McMillan17} potential provides a more
comprehensive model of the Milky Way's disk than either the default or
scaled \texttt{MWPotential2014} potentials.  This difference could be
important for the relatively nearby clusters that dominate our input
sample.  For all of these reasons, the \texttt{McMillan17} potential
likely provides a better approximation to the true Milky Way potential
than the default and scaled \texttt{MWPotential2014} potentials.

Our orbit integrations produce 100 plausible orbits for each of
the 57 globular clusters listed in Table \ref{tab:input} in the
default \texttt{MWPotential2014}, scaled \texttt{MWPotential2014}, and
\texttt{McMillan17} potentials.  We calculate the 16th, 50th, and 84th
percentiles of the specific orbital energy (SOE) distribution for every
cluster and use those data to define each cluster's specific orbital
energy and uncertainty in each potential.  We list the specific
orbital energies derived in this way for each cluster in Table
\ref{tab:inference}.

We use the \texttt{statsmodels} python module \citep{Seabold_2010} to fit
an ordinary least squares linear regression of the form $\mathrm{SOE}
= \beta_0 + \beta_{\tau} \tau + \beta_{M} \mathrm{[M/H]}$ to the
age--metallicity--specific orbital energy distribution produced by
each set of initial conditions in all three potentials.  We aggregate
the results and calculate the 16th, 50th, and 84 percentiles of each
distribution.  We report the results in Table \ref{tab:model}.  In all
three potentials, we find evidence of an age--metallicity--specific
orbital energy relationship in which relatively old, metal-rich globular
clusters have more negative specific orbital energies and are therefore
more tightly bound to the Milky Way than relatively metal-poor,
young globular clusters.  This relationship is especially strong in
the \texttt{McMillan17} potential, as $\beta_{\tau}$ and $\beta_{M}$
in that potential are significant at approximately the 5-$\sigma$ level.

We separately study the age--metallicity--specific orbital energy
relationship for metal-rich and metal-poor clusters.  We split the
complete sample into metal-rich/metal-poor subsamples at $[\text{M/H}]
= -0.8$ because that cleanly separates the two peaks of the complete
sample's bimodal metallicity distribution.  We find similar results
for both the complete sample and the metal-poor subsample of globular
clusters that we define as the 45 globular clusters with $[\text{M/H}]
\leq -0.8$.  The $t$ values and therefore the statistical significances
of the coefficients defining the relation are generally larger in
absolute value for the metal-poor subsample.  For the 12 metal-rich
clusters with $[\text{M/H}] > -0.8$, the age--metallicity--specific
orbital energy relationship is much weaker.  Indeed, if we exclude the
two metal-rich globular clusters associated with the Sagittarius dSph,
then in both the default and scaled \texttt{MWPotential2014} potentials
the coefficients defining the age--metallicity--specific orbital relation
energy change signs such that older, more metal-rich clusters are less
tightly bound to the Milky Way.  In this case, there is no significant
age--metallicity--specific orbital energy relationship for the the
\texttt{McMillan17} potential.

\begin{deluxetable*}{lccccccc}
\tablenum{3}
\tablecaption{Median Regression Model Coefficients and $t$ Values}
\tablehead{
\colhead{Sample} &
\colhead{$\beta_0$}&
\colhead{$\beta_0$} &
\colhead{$\beta_{\tau}$} &
\colhead{$\beta_{\tau}$} &
\colhead{$\beta_M$} &
\colhead{$\beta_M$} &
\colhead{$R^2$} \\
\colhead{} &
\colhead{($10^4$)} &
\colhead{$t$ Value} &
\colhead{($10^4$)} &
\colhead{$t$ Value} &
\colhead{($10^4$)} &
\colhead{$t$ Value} &
\colhead{}}
\startdata
Default \texttt{MWPotential2014} & & & & & & \\
Complete & -5.39 $\pm$ 2.32 & -2.33 & -4.12 $\pm$ 2.54 & -1.62 & -1.16 $\pm$ 0.63 & -1.84 & 0.081 \\
Metal-poor & -2.21 $\pm$ 3.07 & -0.72 & -8.64 $\pm$ 3.62 & -2.39 & -1.98 $\pm$ 0.92 & -2.16 & 0.146 \\
Metal-rich\tablenotemark{a} & -11.26 $\pm$ 5.19 & -2.17 & -1.38 $\pm$ 4.20 & -0.33 & -7.45 $\pm$ 8.09 & -0.92 & 0.087 \\
Metal-rich\tablenotemark{b} & -22.52 $\pm$ 10.02 & -2.25 & 19.30 $\pm$ 10.80 & 1.79 & 9.34 $\pm$ 10.16 & 0.92 & 0.317 \\
\hline
Scaled \texttt{MWPotential2014} & & & & & & \\
Complete & -10.17 $\pm$ 2.44 & -4.16 & -3.95 $\pm$ 2.68 & -1.47 & -1.40 $\pm$ 0.67 & -2.10 & 0.089 \\
Metal-poor & -5.89 $\pm$ 3.19 & -1.85 & -9.53 $\pm$ 3.76 & -2.54 & -2.15 $\pm$ 0.95 & -2.26 & 0.160 \\
Metal-rich\tablenotemark{a} & -18.20 $\pm$ 5.42 & -3.36 & -0.12 $\pm$ 4.38 & -0.03 & -9.53 $\pm$ 8.46 & -1.13 & 0.131 \\
Metal-rich\tablenotemark{b} & -30.52 $\pm$ 9.72 & -3.14 & 22.91 $\pm$ 10.48 & 2.19 & 9.52 $\pm$ 9.85 & 0.97 & 0.407 \\
\hline
\texttt{McMillan17} & & & & & & \\
Complete & 1.39 $\pm$ 3.95 & 0.35 & -24.52 $\pm$ 4.33 & -5.66 & -4.85 $\pm$ 1.08 & -4.49 & 0.435 \\
Metal-poor & 2.80 $\pm$ 5.23 & 0.54 & -28.18 $\pm$ 6.17 & -4.57 & -6.32 $\pm$ 1.56 & -4.05 & 0.381 \\
Metal-rich\tablenotemark{a} & -3.86 $\pm$ 9.24 & -0.42 & -24.76 $\pm$ 7.47 & -3.32 & -15.79 $\pm$ 14.41 & -1.10 & 0.551 \\
Metal-rich\tablenotemark{b} & -38.61 $\pm$ 19.16 & -2.02 & 15.39 $\pm$ 20.67 & 0.745 & -5.92 $\pm$ 19.43 & -0.31 & 0.123
\enddata
\tablecomments{Median age--metallicity--specific orbital energy
model coefficients, coefficient uncertainties, $t$ values, and $R^2$
values calculated from 100 Monte Carlo iterations for the complete,
metal-poor, and metal-rich subsamples (both including and excluding
clusters associated with the Sagittarius dSph) assuming all three
potentials.\label{tab:model}}
\tablenotetext{a}{All metal-rich globular clusters}
\tablenotetext{b}{Metal-rich globular clusters excluding those associated
with the Sagittarius dSph}
\end{deluxetable*}

We use $F$ tests to compare the full age--metallicity--specific orbital
energy model we derived above with each of its nested submodels to verify
that the full age--metallicity--specific orbital energy model best
describes the data in Table \ref{tab:inference}.  For each subsample
in each potential, we tested three null hypotheses: $\beta_{\tau}=0$,
$\beta_M=0$, and $\beta_{\tau}=\beta_M=0$.  We report the results of
those model comparisons in Table \ref{tab:modelcompare1}.  For the
metal-poor subsample in all three potentials, the $p$ values for
each test are small enough to confidently reject the null hypotheses.
The implication is that the full linear model is a better description of
the metal-poor subsample than any of its nested submodels.  In the case
of the \texttt{McMillan17} potential, the full linear model is a better
description of the complete sample than any of its nested submodels.
On the other hand, for the metal-rich subsample there is no reason to
prefer the full age--metallicity--specific orbital energy model over any
of its nested submodels.  We also use the Bayesian information criterion
(BIC) to compare models.  Smaller BIC values indicate better models,
and the results in Table \ref{tab:modelcompare2} show that the full
age--metallicity--specific orbital energy model is preferred to any of its
nested submodels for both the complete sample and metal-poor subsample
in the \texttt{McMillan17} potential.  All models are approximately
similar in the default and scaled \texttt{MWPotential2014} potentials.

\begin{deluxetable*}{lcccccc}
\tablenum{4}
\tablecaption{Median Model Comparison $F$ Statistics}
\tablehead{
\colhead{Sample} &
\twocolhead{$H_0:\beta_{\tau} = 0$} &
\twocolhead{$H_0:\beta_M = 0$} &
\twocolhead{$H_0:\beta_{\tau} = \beta_M = 0$} \\
\colhead{} &
\colhead{$F$} &
\colhead{$p$} &
\colhead{$F$} &
\colhead{$p$} &
\colhead{$F$} &
\colhead{$p$}}
\startdata
Default \texttt{MWPotential2014} & & & & & & \\
Complete & 2.63 & $1.1\times10^{-1}$ & 3.27 & $7.2\times10^{-2}$ & 2.38 & $1.0\times10^{-1}$ \\
Metal-poor & 5.71 & $2.1\times10^{-2}$ & 4.66 & $3.7\times10^{-2}$ & 3.58 & $3.7\times10^{-2}$ \\
Metal-rich\tablenotemark{a} & 0.11 & $7.5\times10^{-1}$ & 0.85 & $3.8\times10^{-1}$ & 0.43 & $6.6\times10^{-1}$ \\
Metal-rich\tablenotemark{b} & 3.19 & $1.1\times10^{-1}$ & 0.85 & $3.9\times10^{-1}$ & 1.62 & $2.6\times10^{-1}$ \\
\hline
Scaled \texttt{MWPotential2014} & & & & & & \\
Complete & 2.17 & $1.5\times10^{-1}$ & 4.40 & $4.1\times10^{-2}$ & 2.65 & $8.0\times10^{-2}$\\
Metal-poor & 6.44 & $1.5\times10^{-2}$ & 5.12 & $2.9\times10^{-2}$ & 4.00 & $2.6\times10^{-2}$\\
Metal-rich\tablenotemark{a} & 0.00 & $9.8\times10^{-1}$ & 1.27 & $2.9\times10^{-1}$ & 0.68 & $5.3\times10^{-1}$\\
Metal-rich\tablenotemark{b} & 4.78 & $6.5\times10^{-1}$ & 0.93 & $3.7\times10^{-1}$ & 2.40 & $1.6\times10^{-1}$\\
\hline
\texttt{McMillan17} & & & & & & \\
Complete & 32.05 & $5.9\times10^{-7}$ & 20.13 & $3.8\times10^{-5}$ & 20.81 & $2.0\times10^{-7}$ \\
Metal-poor & 20.89 & $4.2\times10^{-5}$ & 16.38 & $2.2\times10^{-4}$ & 12.90 & $4.3\times10^{-5}$ \\
Metal-rich\tablenotemark{a} & 11.00 & $9.0\times10^{-3}$ & 1.20 & $3.0\times10^{-1}$ & 5.52 & $3.0\times10^{-2}$ \\
Metal-rich\tablenotemark{b} & 0.55 & $4.8\times10^{-1}$ & 0.09 & $7.7\times10^{-1}$ & 0.49 & $6.3\times10^{-1}$
\enddata
\tablecomments{$F$ statistics and associated $p$ values comparing
the full three parameter age--metallicity--specific orbital energy
relation derived for each potential to all possible lower-dimensional
submodels.  The large $F$ statistics and small $p$ values indicate
that the full age--metallicity--specific orbital energy relation is
superior to all lower-dimensional nested models for the metal-poor
subsample.\label{tab:modelcompare1}}
\tablenotetext{a}{All metal-rich globular clusters}
\tablenotetext{b}{Metal-rich globular clusters excluding those associated
with the Sagittarius dSph}
\end{deluxetable*}

\begin{deluxetable}{lcccc}
\tablenum{5}
\tabletypesize{\footnotesize}
\tablecaption{Median Model Comparison BIC Statistics}
\tablehead{
\colhead{Sample} &
\colhead{Full Model} &
\colhead{$\beta_{\tau} = 0$} &
\colhead{$\beta_M = 0$} &
\colhead{$\beta_{\tau} = \beta_M = 0$}
}
\startdata
Default \texttt{MWPotential2014} & & & & \\
Complete & 241 & 240 & 240 & 238 \\
Metal-poor & 188 & 190 & 189 & 188 \\
Metal-rich\tablenotemark{a} & 55 & 53 & 54 & 52 \\
Metal-rich\tablenotemark{b} & 43 & 44 & 41 & 42 \\
\hline
Scaled \texttt{MWPotential2014} & & & & \\
Complete & 247 & 245 & 248 & 244 \\
Metal-poor & 192 & 194 & 193 & 192 \\
Metal-rich\tablenotemark{a} & 57 & 54 & 56 & 53 \\
Metal-rich\tablenotemark{b} & 42 & 45 & 41 & 43 \\
\hline
\texttt{McMillan17} & & & & \\
Complete & 302 & 324 & 316 & 326 \\
Metal-poor & 236 & 251 & 247 & 250 \\
Metal-rich\tablenotemark{a} & 69 & 76 & 68 & 74 \\
Metal-rich\tablenotemark{b} & 55 & 54 & 53 & 52
\enddata
\tablecomments{BIC values comparing the full three parameter
age--metallicity--specific orbital energy relation derived for each
potential to all possible lower-dimensional submodels.  The small BIC
values indicate that the full age--metallicity--specific orbital energy
relation is superior to all lower-dimensional nested models for both the
complete sample and its metal-poor subsample in the \texttt{McMillan17}
potential.\label{tab:modelcompare2}}
\tablenotetext{a}{All metal-rich globular clusters}
\tablenotetext{b}{Metal-rich globular clusters excluding those associated
with the Sagittarius dSph}
\end{deluxetable}

\begin{figure*}
\epsscale{0.5}
\plotone{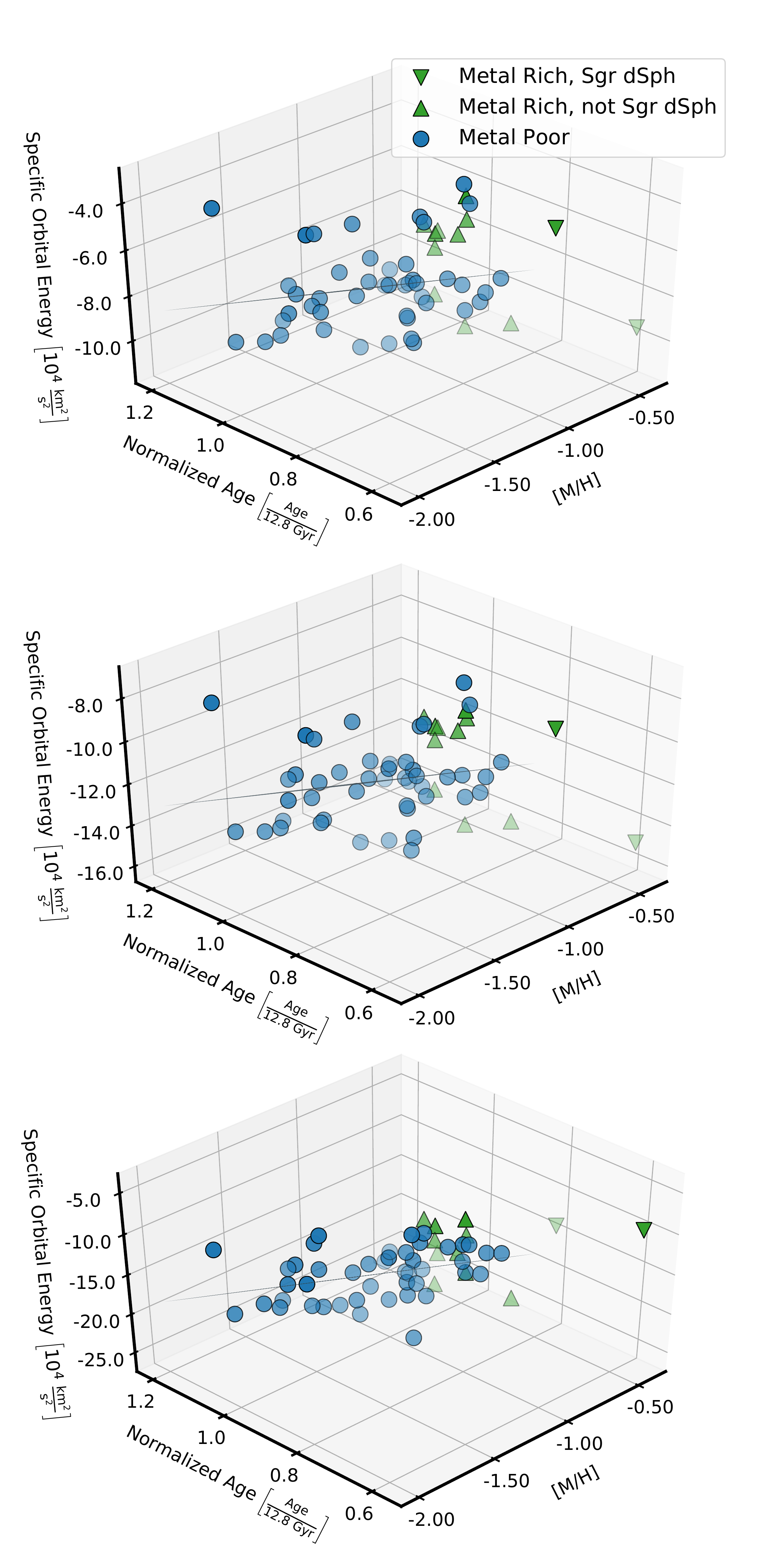}
\caption{Age--metallicity--specific orbital energy distribution of
the Milky Way globular clusters listed in Tables \ref{tab:input} and
\ref{tab:inference}.  We plot metal-poor clusters with $[\text{M/H}] \leq
-0.8$ as blue circles and metal-rich clusters with  $[\text{M/H}] > -0.8$
as green triangles.  Downward-pointing triangles are metal-rich clusters
associated with the Sagittarius dSph galaxy, while upward-pointing
triangles are metal-rich clusters unassociated with the Sagittarius
dSph galaxy.  Each plot is looking down the plane defined by the median
best-fit age--metallicity--specific orbital energy relation for the
metal-poor subsample as reported in Table \ref{tab:model}.  Top: specific
orbital energies calculated assuming the default \texttt{MWPotential2014}
potential from \citet{Bovy_2015}.  Middle: specific orbital
energies calculated assuming the scaled \texttt{MWPotential2014}
potential.  Bottom: specific orbital energies calculated assuming the
\texttt{McMillan17} potential from \citet{McMillan_2017}.  It is visually
clear in the \texttt{McMillan17} potential that the metal-poor subsample
lies in a plane in age--metallicity--specific orbital energy space as
indicated by the highly significant coefficients $\beta_{\tau}$ and
$\beta_{M}$ listed in Table \ref{tab:model} and the small $F$ test $p$
values listed in Table \ref{tab:modelcompare1}.\label{fig:3d}}
\end{figure*}

\begin{figure*}
\plotone{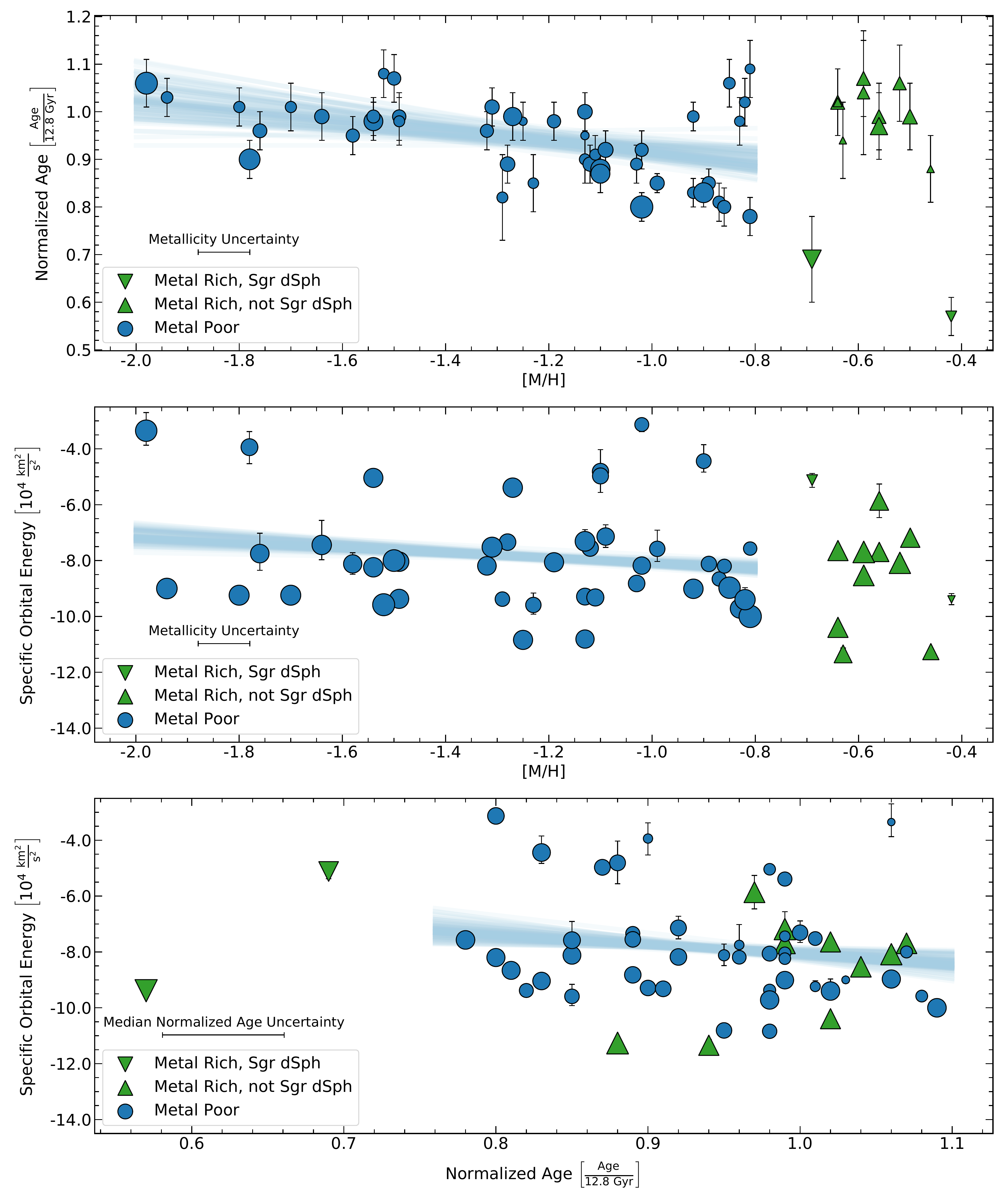}
\caption{Two-dimensional projections of the age--metallicity--specific
orbital energy distribution of the 57 Milky Way globular clusters listed
in Tables \ref{tab:input} and \ref{tab:inference} assuming the default
\texttt{MWPotential2014} potential \citep{Bovy_2015}.  We plot metal-poor
clusters as blue circles and metal-rich clusters as green triangles.
Downward-pointing triangles are metal-rich clusters associated with the
Sagittarius dSph galaxy, while upward-pointing triangles are metal-rich
clusters unassociated with the Sagittarius dSph galaxy.  The sizes of
the points corresponds to the relative values of the suppressed axis.
In blue we indicate the two-dimensional projections of the best-fit
age--metallicity--specific orbital energy relation in the metal-poor
subsample for each of the 100 Monte Carlo trials.  Top: metallicity--age
relation.  Middle: metallicity--specific orbital energy relation.
Bottom: age--specific orbital energy relation.\label{fig:MW2d}}
\end{figure*}

\begin{figure*}
\plotone{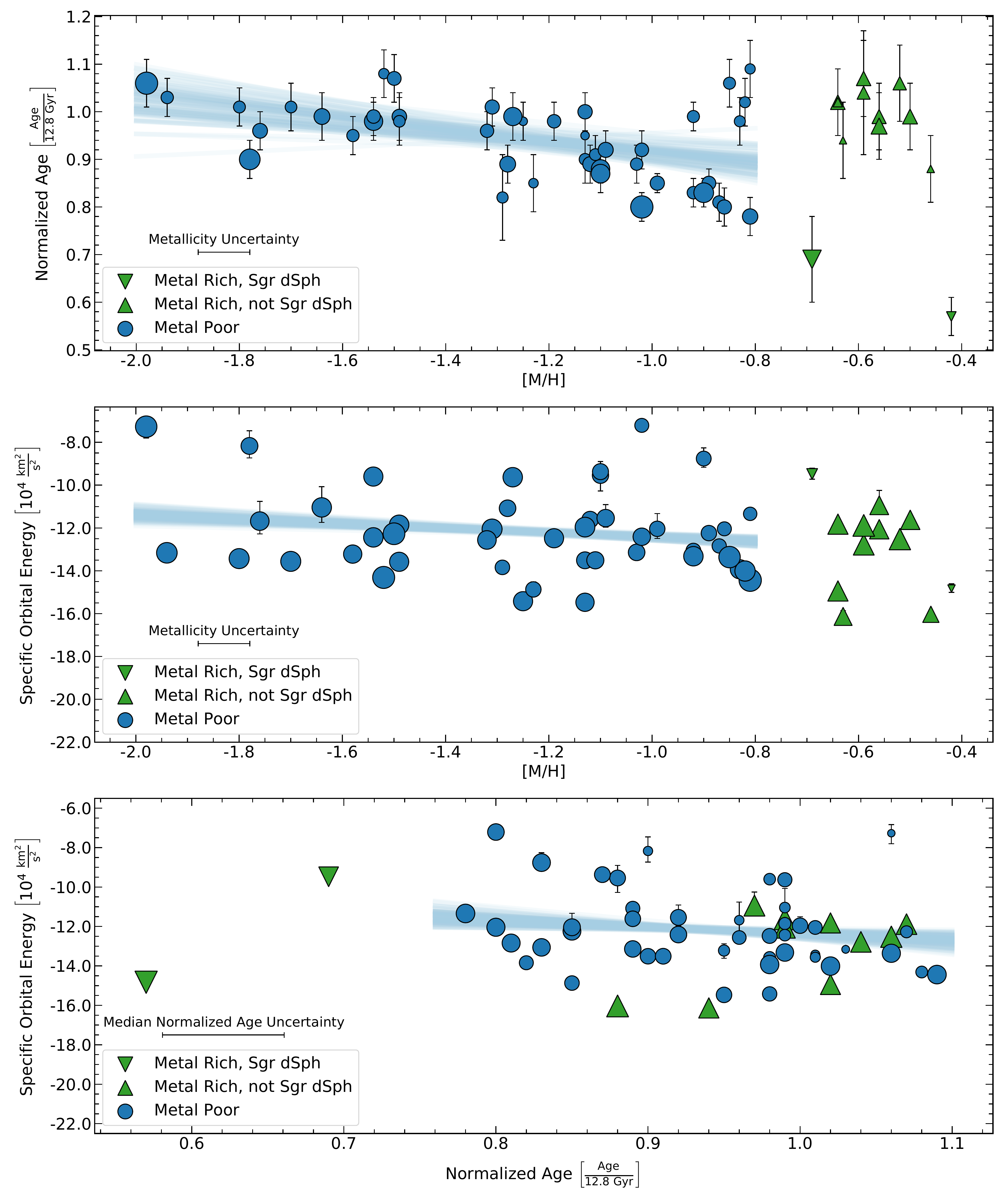}
\caption{Two-dimensional projections of the age--metallicity--specific
orbital energy distribution of the 57 Milky Way globular clusters
listed in Tables \ref{tab:input} and \ref{tab:inference} assuming
the scaled \texttt{MWPotential2014} potential.  We plot metal-poor
clusters as blue circles and metal-rich clusters as green triangles.
Downward-pointing triangles are metal-rich clusters associated with the
Sagittarius dSph galaxy, while upward-pointing triangles are metal-rich
clusters unassociated with the Sagittarius dSph galaxy.  The sizes of
the points corresponds to the relative values of the suppressed axis.
In blue we indicate the two-dimensional projections of the best-fit
age--metallicity--specific orbital energy relation in the metal-poor
subsample for each of the 100 Monte Carlo trials.  Top: metallicity--age
relation.  Middle: metallicity--specific orbital energy relation.
Bottom: age--specific orbital energy relation.\label{fig:SMW2d}}
\end{figure*}

\begin{figure*}
\plotone{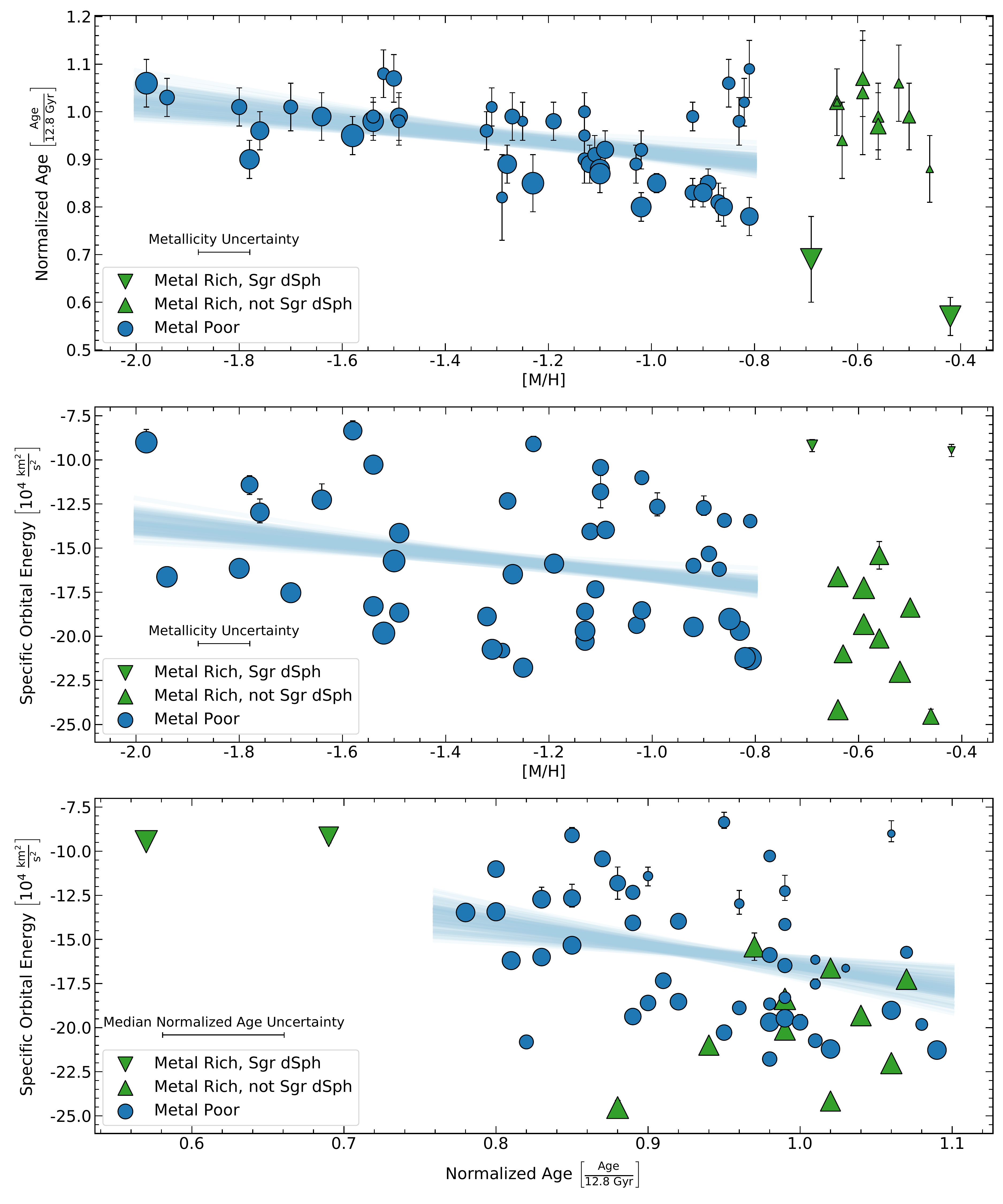}
\caption{Two-dimensional projections of the age--metallicity--specific
orbital energy distribution of the 57 Milky Way globular clusters
listed in Tables \ref{tab:input} and \ref{tab:inference} assuming
\texttt{McMillan17} potential \citep{McMillan_2017}.  We plot metal-poor
clusters as blue circles and metal-rich clusters as green triangles.
Downward-pointing triangles are metal-rich clusters associated with the
Sagittarius dSph galaxy, while upward-pointing triangles are metal-rich
clusters unassociated with the Sagittarius dSph galaxy.  The sizes of
the points corresponds to the relative values of the suppressed axis.
In blue we indicate the two-dimensional projections of the best-fit
age--metallicity--specific orbital energy relation in the metal-poor
subsample for each of the 100 Monte Carlo trials.  Top: metallicity--age
relation.  Middle: metallicity--specific orbital energy relation.
Bottom: age--specific orbital energy relation.\label{fig:Mc2d}}
\end{figure*}

We plot the distribution of the Milky Way's globular cluster systems
in the age--metallicity--specific orbital energy space for all three
potentials in Figure \ref{fig:3d}.  It is clear that the 57 globular
clusters listed in Tables \ref{tab:input} and \ref{tab:inference}
lie in a plane in age--metallicity--specific orbital energy
space in the \texttt{McMillan17} potential, exactly as predicted
by our linear modeling.  We plot two-dimensional projections of
age--metallicity--specific orbital energy space for the default
\texttt{MWPotential2014}, scaled \texttt{MWPotential2014}, and
\texttt{McMillan17} potentials in Figures \ref{fig:MW2d}, \ref{fig:SMW2d},
and \ref{fig:Mc2d}.  There remains significant dispersion about the
linear model visualized in Figures \ref{fig:MW2d}, \ref{fig:SMW2d},
and \ref{fig:Mc2d}, so the age--metallicity--specific orbital energy
relation inference we described in Section 1 does not fully explain the
highly nonlinear process of globular cluster formation.  Nevertheless, the
meaningfully non-zero $R^2$ values presented in Table \ref{tab:model} show
that a simple three-parameter linear model explains at least 15\%---and
possibly as much as 38\%---of the variance in the metal-poor globular
cluster age--metallicity--specific orbital energy relation.

\section{Discussion}\label{discussion}

We find a significant age--metallicity--specific orbital energy
relationship for the metal-poor subsample of 45 globular clusters
with $[\text{M/H}] \leq -0.8$ listed in Tables \ref{tab:input}
and \ref{tab:inference}.  In particular, old, metal-rich globular
clusters tend to have more negative specific orbital energies
and are therefore more tightly bound to the Milky Way than young,
metal-poor globular clusters.  The metal-poor subsample prefers the
full age--metallicity--specific orbital energy relation over any of its
nested submodels.  These results are independent of assumed potential,
although the relationship is much stronger in the \texttt{McMillan17}
potential from \citet{McMillan_2017} than either the default or scaled
\texttt{MWPotential2014} potentials from \citet{Bovy_2015}.  We find
no clear age--metallicity--specific orbital energy relation in the
metal-rich subsample of 10 globular clusters with $[\text{M/H}] > -0.8$
that are not associated with the Sagittarius dSph galaxy.  We also find
in this metal-rich subsample that an age--metallicity--specific orbital
energy relation is not clearly preferred over lower-dimensional models
relating age, metallicity, or specific orbital energy.

As we argued in Section \ref{intro}, accreted metal-rich globular
clusters can only be formed in massive dSph galaxies that experience
stronger dynamical friction and form their first stars more quickly
than lower-mass dSph galaxies.  These two properties imply that in
a globular cluster system with many accreted globular clusters, age
and metallicity should be correlated with each other but inversely
correlated with specific orbital energy.  Our observation that old,
metal-rich globular clusters are more tightly bound to the Milky
Way than young, metal-poor globular clusters confirms the importance
of accretion for the formation of the Milky Way's globular cluster
system.  Our inference also supports previous proposals that a large
fraction of the Milky Way's metal-poor globular clusters were accreted
\citep[e.g.,][]{Forbes_2010,Forbes_2020,Kruijssen_2019b,Kruijssen_2019c,Kruijssen_2020,Massari_2019,Trujillo-Gomez_2021}.

All but two of the metal rich clusters in our sample have tightly bound
orbits and old ages, broadly consistent with an in situ origin for
metal-rich globular clusters.  These two clusters---Terzan 7 and Pal
12---are both young and on energetic orbits.  They have been securely
associated with the Sagittarius dSph galaxy though, and therefore have
been accreted \citep[e.g.,][]{Law_2010,Sohn_2018}.  Because we do not
observe a strong age--metallicity--specific orbital energy for metal-rich
globular clusters unassociated with the Sagittarius dSph, our inferences
do not support an accretion origin for most metal-rich globular clusters.
As there are only 10 metal-rich clusters unassociated with the Sagittarius
dSph in our sample, we are unable to choose between gas-rich mergers or
formation in gas-rich disks for the origin of the Milky Way's metal-rich
globular clusters.

In our analysis, we used relative ages from \citet{Marin_2009} calculated
assuming the \citet{Carretta_1997} metallicity scale and the theoretical
model grid from \citet{Dotter_2007}.  To ensure that our results are
insensitive to the assumed metallicity scale or theoretical model grid
underlying the \citet{Marin_2009} age inferences, we performed the same
analysis using all of the relative ages from \citet{Marin_2009} calculated
assuming both the \citet{Zinn_1984} and \citet{Carretta_1997} metallicity
scales and all of the \citet{Bertelli_1994}, \citet{Girardi_2000},
\citet{Pietrinferni_2004}, and \citet{Dotter_2007} theoretical model
grids.  In all cases, our analysis produces the same trends with
comparable statistical significance.  This should not be surprising,
as \citet{Marin_2009} found that their relative ages inferences were
insensitive to their choice of metallicity scale or theoretical model
grid.  This consistency supports the robustness of our findings.

As a consequence of the simple physical explanation we outline to explain
the age--metallicity--specific orbital energy relation we observe in the
Milky Way's globular cluster system, we predict that the globular cluster
systems of other $L^{\ast}$ galaxies should evince similar relations.
That is, we predict that other $L^{\ast}$ galaxies in isolation or in
low-density group environments like the Local Group should have globular
cluster systems in which old, metal-rich globular clusters have more
negative specific orbital energies and are therefore more tightly bound to
their host halo than young, metal-poor globular clusters.  We argue that
this is a natural outcome of galaxy formation in a $\Lambda$CDM universe.

While we infer the importance of accretion for the origin of
the Milky Way's globular cluster system from an analysis of its
ensemble properties, our approach cannot associate individual
globular clusters with individual accretion events.  Detailed
simulations of Milky Way analogs have shown how a globular cluster
systems properties can be used to reveal that halo's accretion history
\citep[e.g.,][]{Kruijssen_2019b,Kruijssen_2019c,Kruijssen_2020,Pfeffer_2020,Trujillo-Gomez_2021}.
When applied to the Milky Way, that formalism can quantify the properties
of massive satellite galaxies that have since merged with the Milky Way
\citep[e.g.,][]{Myeong_2018,Hughes_2019,Kruijssen_2019c,Kruijssen_2020,Pfeffer_2020,Trujillo-Gomez_2021}.
Moreover, efforts have been made to match the kinematics of clusters to
stellar streams associated with known accretion events.  For example,
\citet{Massari_2019} employed this technique in combination with
in-group age--metallicity relations to assign candidate accreted globular
clusters to specific progenitors.  However, our analysis shows that it is
beneficial to use similar techniques on the entire cluster population,
thereby capturing the general imprint that accretion has had on the
Milky Way's globular cluster system's age--metallicity--specific orbital
energy relation.

Because proper motions are more difficult to infer for distant clusters,
our input sample lacks outer halo clusters.  This bias is unlikely to
affect our conclusion, as the only plausible explanation for weakly
bound outer halo clusters is accretion from dwarf galaxies.  Our input
sample also lacks highly extincted disk and bulge clusters.  This bias
is unlikely to affect our conclusion either, as disk/bulge clusters are
metal-rich and would not appear in our metal-poor subsample even if they
were present in our input sample.

The age--metallicity--specific orbital energy relation we observe in
our metal-poor subsample of globular clusters is much stronger in the
\texttt{McMillan17} potential than in the \texttt{MWPotential2014}
potential.  As discussed in Section \ref{analysis}, the default
\texttt{MWPotential2014} potential represents a comparatively
smaller, lower-mass Milky Way than the \texttt{McMillan17} potential.
In addition, the more massive \texttt{McMillan17} potential appears
to be a better match to Gaia DR2-informed Milky Way virial mass
inferences.  Under the default \texttt{MWPotential2014} potential,
most of the weakly bound clusters are relatively less bound than they
are in the \texttt{McMillan17} potential, though this difference is
diminished in the scaled \texttt{MWPotential2014}.  This results in
a relatively larger spread in orbital energy, which might explain the
weaker age--metallicity--specific orbital energy relation.  Nevertheless,
the age--metallicity--specific orbital energy relation has the same form
in all three potentials and we are confident its existence is robust to
the choice of potential.

\section{Conclusion}\label{conclusion}

We find that in the metal-poor subsample of 45 Milky Way globular
clusters with $[\text{M/H}] \leq -0.8$ in Tables \ref{tab:input} and
\ref{tab:inference}, relatively young or metal-poor globular clusters
are weakly bound to the Milky Way, while relatively old or metal-rich
globular clusters are tightly bound to the Galaxy.  We argue that this
relationship is naturally explained by the accretion of globular clusters
from now-disrupted dwarf galaxies.  We propose that this observation is
a consequence of the combined effects of the stellar mass--metallicity
relationship for dwarf galaxies, the dependence of dynamical friction
on mass, and the relationship between a galaxy's dynamical time and its
central density.  Accreted metal-rich globular clusters can only form
in massive dSph galaxies that can attain high metallicities and will be
strongly affected by dynamical friction.  These more massive dSph galaxies
capable of forming metal-rich clusters will also produce their oldest
stellar populations before lower-mass dSph galaxies that form in lower
$\sigma$ peaks in the universe's primordial matter density distribution.
These lower-mass dSph galaxies will be much less affected by dynamical
friction and can only produce low-metallicity globular clusters.
Because these properties are independent of the specific accretion
history of the Milky Way, we assert that they are a natural outcome
of galaxy formation in a $\Lambda$CDM universe.  We predict that the
globular cluster systems of other $L^{\ast}$ galaxies in isolation
or in low-density groups like the Local Group will lie in a plane in
age--metallicity--specific orbital energy space.

\acknowledgments
We thank Brendan Griffen and David Nataf for insightful suggestions that
improved our analyses.  This work has made use of data from the European
Space Agency (ESA) mission Gaia (\url{https://www.cosmos.esa.int/gaia}),
processed by the Gaia Data Processing and Analysis Consortium (DPAC,
\url{https://www.cosmos.esa.int/web/gaia/dpac/consortium}).  Funding for
the DPAC has been provided by national institutions, in particular the
institutions participating in the Gaia Multilateral Agreement.  This
research has made use of NASA's Astrophysics Data System Bibliographic
Services.  This research has made use of the SIMBAD database, operated
at CDS, Strasbourg, France \citep{Wenger_2000}.  This research has
made use of the VizieR catalogue access tool, CDS, Strasbourg, France.
The original description of the VizieR service was published in A\&AS 143,
23 \citep{Ochsenbein_2000}.

\software{\texttt{galpy} \citep{Bovy_2015},
          \texttt{numpy} \citep{vanderWalt_2011},
          \texttt{pandas} \citep{McKinney_2010},
          \texttt{statsmodels} \citep{Seabold_2010}
          }

\clearpage
\bibliography{ms}{}

\begin{thebibliography}{}
\expandafter\ifx\csname natexlab\endcsname\relax\def\natexlab#1{#1}\fi

\bibitem[{{Arenou} {et~al.}(2018){Arenou}, {Luri}, {Babusiaux}, {Fabricius},
  {Helmi}, {Muraveva}, {Robin}, {Spoto}, {Vallenari}, {Antoja},
  {Cantat-Gaudin}, {Jordi}, {Leclerc}, {Reyl{\'e}}, {Romero-G{\'o}mez}, {Shih},
  {Soria}, {Barache}, {Bossini}, {Bragaglia}, {Breddels}, {Fabrizio},
  {Lambert}, {Marrese}, {Massari}, {Moitinho}, {Robichon}, {Ruiz-Dern},
  {Sordo}, {Veljanoski}, {Eyer}, {Jasniewicz}, {Pancino}, {Soubiran}, {Spagna},
  {Tanga}, {Turon}, \& {Zurbach}}]{Arenou_2018}
{Arenou}, F., {Luri}, X., {Babusiaux}, C., {et~al.} 2018, \aap, 616, A17

\bibitem[{{Ashman} \& {Zepf}(1992)}]{Ashman_1992}
{Ashman}, K.~M., \& {Zepf}, S.~E. 1992, \apj, 384, 50

\bibitem[{{Baumgardt} {et~al.}(2019){Baumgardt}, {Hilker}, {Sollima}, \&
  {Bellini}}]{Baumgardt_2019}
{Baumgardt}, H., {Hilker}, M., {Sollima}, A., \& {Bellini}, A. 2019, \mnras,
  482, 5138

\bibitem[{{Bertelli} {et~al.}(1994){Bertelli}, {Bressan}, {Chiosi}, {Fagotto},
  \& {Nasi}}]{Bertelli_1994}
{Bertelli}, G., {Bressan}, A., {Chiosi}, C., {Fagotto}, F., \& {Nasi}, E. 1994,
  \aaps, 106, 275

\bibitem[{{Binney} \& {Tremaine}(2008)}]{Binney_2008}
{Binney}, J., \& {Tremaine}, S. 2008, {Galactic Dynamics: Second Edition}

\bibitem[{{Bjork} \& {Chaboyer}(2006)}]{Bjork_2006}
{Bjork}, S.~R., \& {Chaboyer}, B. 2006, \apj, 641, 1102

\bibitem[{{Bland-Hawthorn} \& {Gerhard}(2016)}]{Bland_2016}
{Bland-Hawthorn}, J., \& {Gerhard}, O. 2016, \araa, 54, 529

\bibitem[{{Bovy}(2015)}]{Bovy_2015}
{Bovy}, J. 2015, \apjs, 216, 29

\bibitem[{{Buonanno} {et~al.}(1999){Buonanno}, {Corsi}, {Castellani},
  {Marconi}, {Fusi Pecci}, \& {Zinn}}]{Buonanno_1999}
{Buonanno}, R., {Corsi}, C.~E., {Castellani}, M., {et~al.} 1999, \aj, 118, 1671

\bibitem[{{Buonanno} {et~al.}(1998){Buonanno}, {Corsi}, {Zinn}, {Pecci},
  {Hardy}, \& {Suntzeff}}]{Buonanno_1998}
{Buonanno}, R., {Corsi}, C.~E., {Zinn}, R., {et~al.} 1998, \apjl, 501, L33

\bibitem[{{Carretta} {et~al.}(2014){Carretta}, {Bragaglia}, {Gratton},
  {D'Orazi}, {Lucatello}, \& {Sollima}}]{Carretta_2014}
{Carretta}, E., {Bragaglia}, A., {Gratton}, R.~G., {et~al.} 2014, \aap, 561,
  A87

\bibitem[{{Carretta} \& {Gratton}(1997)}]{Carretta_1997}
{Carretta}, E., \& {Gratton}, R.~G. 1997, \aaps, 121, 95

\bibitem[{{Carretta} {et~al.}(2010){Carretta}, {Bragaglia}, {Gratton},
  {Lucatello}, {Bellazzini}, {Catanzaro}, {Leone}, {Momany}, {Piotto}, \&
  {D'Orazi}}]{Carretta_2010}
{Carretta}, E., {Bragaglia}, A., {Gratton}, R.~G., {et~al.} 2010, \aap, 520,
  A95

\bibitem[{{Chaboyer} {et~al.}(2001){Chaboyer}, {Fenton}, {Nelan}, {Patnaude},
  \& {Simon}}]{Chaboyer_2001}
{Chaboyer}, B., {Fenton}, W.~H., {Nelan}, J.~E., {Patnaude}, D.~J., \& {Simon},
  F.~E. 2001, \apj, 562, 521

\bibitem[{{Choksi} {et~al.}(2018){Choksi}, {Gnedin}, \& {Li}}]{Choksi_2018}
{Choksi}, N., {Gnedin}, O.~Y., \& {Li}, H. 2018, \mnras, 480, 2343

\bibitem[{{Chou} {et~al.}(2007){Chou}, {Majewski}, {Cunha}, {Smith},
  {Patterson}, {Mart{\'\i}nez-Delgado}, {Law}, {Crane}, {Mu{\~n}oz}, {Garcia
  L{\'o}pez}, {Geisler}, \& {Skrutskie}}]{Chou_2007}
{Chou}, M.-Y., {Majewski}, S.~R., {Cunha}, K., {et~al.} 2007, \apj, 670, 346

\bibitem[{{Cohen}(2004)}]{Cohen_2004}
{Cohen}, J.~G. 2004, \aj, 127, 1545

\bibitem[{{Crowley} {et~al.}(2016){Crowley}, {Kohley}, {Hambly}, {Davidson},
  {Abreu}, {van Leeuwen}, {Fabricius}, {Seabroke}, {de Bruijne}, {Short},
  {Lindegren}, {Brown}, {Sarri}, {Gare}, {Prusti}, {Prod'homme}, {Mora},
  {Mart{\'\i}n-Fleitas}, {Raison}, {Lammers}, {O'Mullane}, \&
  {Jansen}}]{Crowley_2016}
{Crowley}, C., {Kohley}, R., {Hambly}, N.~C., {et~al.} 2016, \aap, 595, A6

\bibitem[{{Dotter} {et~al.}(2007){Dotter}, {Chaboyer}, {Jevremovi{\'c}},
  {Baron}, {Ferguson}, {Sarajedini}, \& {Anderson}}]{Dotter_2007}
{Dotter}, A., {Chaboyer}, B., {Jevremovi{\'c}}, D., {et~al.} 2007, \aj, 134,
  376

\bibitem[{Eadie \& Juri{\'{c}}(2019)}]{Eadie_2019}
Eadie, G., \& Juri{\'{c}}, M. 2019, The Astrophysical Journal, 875, 159

\bibitem[{{Erkal} {et~al.}(2019){Erkal}, {Belokurov}, {Laporte}, {Koposov},
  {Li}, {Grillmair}, {Kallivayalil}, {Price-Whelan}, {Evans}, {Hawkins},
  {Hendel}, {Mateu}, {Navarro}, {del Pino}, {Slater}, {Sohn}, \& {Orphan Aspen
  Treasury Collaboration}}]{Erkal_2019}
{Erkal}, D., {Belokurov}, V., {Laporte}, C.~F.~P., {et~al.} 2019, \mnras, 487,
  2685

\bibitem[{{Fabricius} {et~al.}(2016){Fabricius}, {Bastian}, {Portell},
  {Casta{\~n}eda}, {Davidson}, {Hambly}, {Clotet}, {Biermann}, {Mora},
  {Busonero}, {Riva}, {Brown}, {Smart}, {Lammers}, {Torra}, {Drimmel},
  {Gracia}, {L{\"o}ffler}, {Spagna}, {Lindegren}, {Klioner}, {Andrei}, {Bach},
  {Bramante}, {Br{\"u}semeister}, {Busso}, {Carrasco}, {Gai}, {Garralda},
  {Gonz{\'a}lez-Vidal}, {Guerra}, {Hauser}, {Jordan}, {Jordi}, {Lenhardt},
  {Mignard}, {Messineo}, {Mulone}, {Serraller}, {Stampa}, {Tanga}, {van
  Elteren}, {van Reeven}, {Voss}, {Abbas}, {Allasia}, {Altmann}, {Anton},
  {Barache}, {Becciani}, {Berthier}, {Bianchi}, {Bombrun}, {Bouquillon},
  {Bourda}, {Bucciarelli}, {Butkevich}, {Buzzi}, {Cancelliere}, {Carlucci},
  {Charlot}, {Collins}, {Comoretto}, {Cross}, {Crosta}, {de Felice}, {Fienga},
  {Figueras}, {Fraile}, {Geyer}, {Hernandez}, {Hobbs}, {Hofmann}, {Liao},
  {Licata}, {Martino}, {McMillan}, {Michalik}, {Morbidelli}, {Parsons},
  {Pecoraro}, {Ramos-Lerate}, {Sarasso}, {Siddiqui}, {Steele},
  {Steidelm{\"u}ller}, {Taris}, {Vecchiato}, {Abreu}, {Anglada}, {Boudreault},
  {Cropper}, {Holl}, {Cheek}, {Crowley}, {Fleitas}, {Hutton}, {Osinde},
  {Rowell}, {Salguero}, {Utrilla}, {Blagorodnova}, {Soffel}, {Osorio},
  {Vicente}, {Cambras}, \& {Bernstein}}]{Fabricius_2016}
{Fabricius}, C., {Bastian}, U., {Portell}, J., {et~al.} 2016, \aap, 595, A3

\bibitem[{{Forbes}(2020)}]{Forbes_2020}
{Forbes}, D.~A. 2020, \mnras, 493, 847

\bibitem[{{Forbes} \& {Bridges}(2010)}]{Forbes_2010}
{Forbes}, D.~A., \& {Bridges}, T. 2010, \mnras, 404, 1203

\bibitem[{{Forbes} {et~al.}(2018){Forbes}, {Bastian}, {Gieles}, {Crain},
  {Kruijssen}, {Larsen}, {Ploeckinger}, {Agertz}, {Trenti}, {Ferguson},
  {Pfeffer}, \& {Gnedin}}]{Forbes_2018}
{Forbes}, D.~A., {Bastian}, N., {Gieles}, M., {et~al.} 2018, Proceedings of the
  Royal Society of London Series A, 474, 20170616

\bibitem[{{Gaia Collaboration} {et~al.}(2016){Gaia Collaboration}, {Prusti},
  {de Bruijne}, {Brown}, {Vallenari}, {Babusiaux}, {Bailer-Jones}, {Bastian},
  {Biermann}, {Evans}, {Eyer}, {Jansen}, {Jordi}, {Klioner}, {Lammers},
  {Lindegren}, {Luri}, {Mignard}, {Milligan}, {Panem}, {Poinsignon},
  {Pourbaix}, {Randich}, {Sarri}, {Sartoretti}, {Siddiqui}, {Soubiran},
  {Valette}, {van Leeuwen}, {Walton}, {Aerts}, {Arenou}, {Cropper}, {Drimmel},
  {H{\o}g}, {Katz}, {Lattanzi}, {O'Mullane}, {Grebel}, {Holland}, {Huc},
  {Passot}, {Bramante}, {Cacciari}, {Casta{\~n}eda}, {Chaoul}, {Cheek}, {De
  Angeli}, {Fabricius}, {Guerra}, {Hern{\'a}ndez}, {Jean-Antoine-Piccolo},
  {Masana}, {Messineo}, {Mowlavi}, {Nienartowicz}, {Ord{\'o}{\~n}ez-Blanco},
  {Panuzzo}, {Portell}, {Richards}, {Riello}, {Seabroke}, {Tanga},
  {Th{\'e}venin}, {Torra}, {Els}, {Gracia-Abril}, {Comoretto},
  {Garcia-Reinaldos}, {Lock}, {Mercier}, {Altmann}, {Andrae}, {Astraatmadja},
  {Bellas-Velidis}, {Benson}, {Berthier}, {Blomme}, {Busso}, {Carry},
  {Cellino}, {Clementini}, {Cowell}, {Creevey}, {Cuypers}, {Davidson}, {De
  Ridder}, {de Torres}, {Delchambre}, {Dell'Oro}, {Ducourant}, {Fr{\'e}mat},
  {Garc{\'\i}a-Torres}, {Gosset}, {Halbwachs}, {Hambly}, {Harrison}, {Hauser},
  {Hestroffer}, {Hodgkin}, {Huckle}, {Hutton}, {Jasniewicz}, {Jordan},
  {Kontizas}, {Korn}, {Lanzafame}, {Manteiga}, {Moitinho}, {Muinonen},
  {Osinde}, {Pancino}, {Pauwels}, {Petit}, {Recio-Blanco}, {Robin}, {Sarro},
  {Siopis}, {Smith}, {Smith}, {Sozzetti}, {Thuillot}, {van Reeven}, {Viala},
  {Abbas}, {Abreu Aramburu}, {Accart}, {Aguado}, {Allan}, {Allasia},
  {Altavilla}, {{\'A}lvarez}, {Alves}, {Anderson}, {Andrei}, {Anglada Varela},
  {Antiche}, {Antoja}, {Ant{\'o}n}, {Arcay}, {Atzei}, {Ayache}, {Bach},
  {Baker}, {Balaguer-N{\'u}{\~n}ez}, {Barache}, {Barata}, {Barbier}, {Barblan},
  {Baroni}, {Barrado y Navascu{\'e}s}, {Barros}, {Barstow}, {Becciani},
  {Bellazzini}, {Bellei}, {Bello Garc{\'\i}a}, {Belokurov}, {Bendjoya},
  {Berihuete}, {Bianchi}, {Bienaym{\'e}}, {Billebaud}, {Blagorodnova},
  {Blanco-Cuaresma}, {Boch}, {Bombrun}, {Borrachero}, {Bouquillon}, {Bourda},
  {Bouy}, {Bragaglia}, {Breddels}, {Brouillet}, {Br{\"u}semeister},
  {Bucciarelli}, {Budnik}, {Burgess}, {Burgon}, {Burlacu}, {Busonero}, {Buzzi},
  {Caffau}, {Cambras}, {Campbell}, {Cancelliere}, {Cantat-Gaudin}, {Carlucci},
  {Carrasco}, {Castellani}, {Charlot}, {Charnas}, {Charvet}, {Chassat},
  {Chiavassa}, {Clotet}, {Cocozza}, {Collins}, {Collins}, {Costigan}, {Crifo},
  {Cross}, {Crosta}, {Crowley}, {Dafonte}, {Damerdji}, {Dapergolas}, {David},
  {David}, {De Cat}, {de Felice}, {de Laverny}, {De Luise}, {De March}, {de
  Martino}, {de Souza}, {Debosscher}, {del Pozo}, {Delbo}, {Delgado},
  {Delgado}, {di Marco}, {Di Matteo}, {Diakite}, {Distefano}, {Dolding}, {Dos
  Anjos}, {Drazinos}, {Dur{\'a}n}, {Dzigan}, {Ecale}, {Edvardsson}, {Enke},
  {Erdmann}, {Escolar}, {Espina}, {Evans}, {Eynard Bontemps}, {Fabre},
  {Fabrizio}, {Faigler}, {Falc{\~a}o}, {Farr{\`a}s Casas}, {Faye}, {Federici},
  {Fedorets}, {Fern{\'a}ndez-Hern{\'a}ndez}, {Fernique}, {Fienga}, {Figueras},
  {Filippi}, {Findeisen}, {Fonti}, {Fouesneau}, {Fraile}, {Fraser}, {Fuchs},
  {Furnell}, {Gai}, {Galleti}, {Galluccio}, {Garabato}, {Garc{\'\i}a-Sedano},
  {Gar{\'e}}, {Garofalo}, {Garralda}, {Gavras}, {Gerssen}, {Geyer}, {Gilmore},
  {Girona}, {Giuffrida}, {Gomes}, {Gonz{\'a}lez-Marcos},
  {Gonz{\'a}lez-N{\'u}{\~n}ez}, {Gonz{\'a}lez-Vidal}, {Granvik}, {Guerrier},
  {Guillout}, {Guiraud}, {G{\'u}rpide}, {Guti{\'e}rrez-S{\'a}nchez}, {Guy},
  {Haigron}, {Hatzidimitriou}, {Haywood}, {Heiter}, {Helmi}, {Hobbs},
  {Hofmann}, {Holl}, {Holland }, {Hunt}, {Hypki}, {Icardi}, {Irwin}, {Jevardat
  de Fombelle}, {Jofr{\'e}}, {Jonker}, {Jorissen}, {Julbe}, {Karampelas},
  {Kochoska}, {Kohley}, {Kolenberg}, {Kontizas}, {Koposov}, {Kordopatis},
  {Koubsky}, {Kowalczyk}, {Krone-Martins}, {Kudryashova}, {Kull}, {Bachchan},
  {Lacoste-Seris}, {Lanza}, {Lavigne}, {Le Poncin-Lafitte}, {Lebreton},
  {Lebzelter}, {Leccia}, {Leclerc}, {Lecoeur-Taibi}, {Lemaitre}, {Lenhardt},
  {Leroux}, {Liao}, {Licata}, {Lindstr{\o}m}, {Lister}, {Livanou}, {Lobel},
  {L{\"o}ffler}, {L{\'o}pez}, {Lopez-Lozano}, {Lorenz}, {Loureiro},
  {MacDonald}, {Magalh{\~a}es Fernandes}, {Managau}, {Mann}, {Mantelet},
  {Marchal}, {Marchant}, {Marconi}, {Marie}, {Marinoni}, {Marrese},
  {Marschalk{\'o}}, {Marshall}, {Mart{\'\i}n-Fleitas}, {Martino}, {Mary},
  {Matijevi{\v{c}}}, {Mazeh}, {McMillan}, {Messina}, {Mestre}, {Michalik},
  {Millar}, {Miranda}, {Molina}, {Molinaro}, {Molinaro}, {Moln{\'a}r},
  {Moniez}, {Montegriffo}, {Monteiro}, {Mor}, {Mora}, {Morbidelli}, {Morel},
  {Morgenthaler}, {Morley}, {Morris}, {Mulone}, {Muraveva}, {Musella},
  {Narbonne}, {Nelemans}, {Nicastro}, {Noval}, {Ord{\'e}novic},
  {Ordieres-Mer{\'e}}, {Osborne}, {Pagani}, {Pagano}, {Pailler}, {Palacin},
  {Palaversa}, {Parsons}, {Paulsen}, {Pecoraro}, {Pedrosa}, {Pentik{\"a}inen},
  {Pereira}, {Pichon}, {Piersimoni}, {Pineau}, {Plachy}, {Plum}, {Poujoulet},
  {Pr{\v{s}}a}, {Pulone}, {Ragaini}, {Rago}, {Rambaux}, {Ramos-Lerate},
  {Ranalli}, {Rauw}, {Read}, {Regibo}, {Renk}, {Reyl{\'e}}, {Ribeiro},
  {Rimoldini}, {Ripepi}, {Riva}, {Rixon}, {Roelens}, {Romero-G{\'o}mez},
  {Rowell}, {Royer}, {Rudolph}, {Ruiz-Dern}, {Sadowski}, {Sagrist{\`a}
  Sell{\'e}s}, {Sahlmann}, {Salgado}, {Salguero}, {Sarasso}, {Savietto},
  {Schnorhk}, {Schultheis}, {Sciacca}, {Segol}, {Segovia}, {Segransan},
  {Serpell}, {Shih}, {Smareglia}, {Smart}, {Smith}, {Solano}, {Solitro},
  {Sordo}, {Soria Nieto}, {Souchay}, {Spagna}, {Spoto}, {Stampa}, {Steele},
  {Steidelm{\"u}ller}, {Stephenson}, {Stoev}, {Suess}, {S{\"u}veges}, {Surdej},
  {Szabados}, {Szegedi-Elek}, {Tapiador}, {Taris}, {Tauran}, {Taylor},
  {Teixeira}, {Terrett}, {Tingley}, {Trager}, {Turon}, {Ulla}, {Utrilla},
  {Valentini}, {van Elteren}, {Van Hemelryck}, {van Leeuwen}, {Varadi},
  {Vecchiato}, {Veljanoski}, {Via}, {Vicente}, {Vogt}, {Voss}, {Votruba},
  {Voutsinas}, {Walmsley}, {Weiler}, {Weingrill}, {Werner}, {Wevers},
  {Whitehead}, {Wyrzykowski}, {Yoldas}, {{\v{Z}}erjal}, {Zucker}, {Zurbach},
  {Zwitter}, {Alecu}, {Allen}, {Allende Prieto}, {Amorim},
  {Anglada-Escud{\'e}}, {Arsenijevic}, {Azaz}, {Balm}, {Beck}, {Bernstein},
  {Bigot}, {Bijaoui}, {Blasco}, {Bonfigli}, {Bono}, {Boudreault}, {Bressan},
  {Brown}, {Brunet}, {Bunclark}, {Buonanno}, {Butkevich}, {Carret}, {Carrion},
  {Chemin}, {Ch{\'e}reau}, {Corcione}, {Darmigny}, {de Boer}, {de Teodoro}, {de
  Zeeuw}, {Delle Luche}, {Domingues}, {Dubath}, {Fodor}, {Fr{\'e}zouls},
  {Fries}, {Fustes}, {Fyfe}, {Gallardo}, {Gallegos}, {Gardiol}, {Gebran},
  {Gomboc}, {G{\'o}mez}, {Grux}, {Gueguen}, {Heyrovsky}, {Hoar}, {Iannicola},
  {Isasi Parache}, {Janotto}, {Joliet}, {Jonckheere}, {Keil}, {Kim},
  {Klagyivik}, {Klar}, {Knude}, {Kochukhov}, {Kolka}, {Kos}, {Kutka}, {Lainey},
  {LeBouquin}, {Liu}, {Loreggia}, {Makarov}, {Marseille}, {Martayan},
  {Martinez-Rubi}, {Massart}, {Meynadier}, {Mignot}, {Munari}, {Nguyen},
  {Nordlander}, {Ocvirk}, {O'Flaherty}, {Olias Sanz}, {Ortiz}, {Osorio},
  {Oszkiewicz}, {Ouzounis}, {Palmer}, {Park}, {Pasquato}, {Peltzer}, {Peralta},
  {P{\'e}turaud}, {Pieniluoma}, {Pigozzi}, {Poels}, {Prat}, {Prod'homme},
  {Raison}, {Rebordao}, {Risquez}, {Rocca-Volmerange}, {Rosen}, {Ruiz-Fuertes},
  {Russo}, {Sembay}, {Serraller Vizcaino}, {Short}, {Siebert}, {Silva},
  {Sinachopoulos}, {Slezak}, {Soffel}, {Sosnowska}, {Strai{\v{z}}ys}, {ter
  Linden}, {Terrell}, {Theil}, {Tiede}, {Troisi}, {Tsalmantza}, {Tur},
  {Vaccari}, {Vachier}, {Valles}, {Van Hamme}, {Veltz}, {Virtanen}, {Wallut},
  {Wichmann}, {Wilkinson}, {Ziaeepour}, \& {Zschocke}}]{Gaia_2016}
{Gaia Collaboration}, {Prusti}, T., {de Bruijne}, J.~H.~J., {et~al.} 2016,
  \aap, 595, A1

\bibitem[{{Gaia Collaboration} {et~al.}(2018{\natexlab{a}}){Gaia
  Collaboration}, {Helmi}, {van Leeuwen}, {McMillan}, {Massari}, {Antoja},
  {Robin}, {Lindegren}, {Bastian}, {Arenou}, {Babusiaux}, {Biermann},
  {Breddels}, {Hobbs}, {Jordi}, {Pancino}, {Reyl{\'e}}, {Veljanoski}, {Brown},
  {Vallenari}, {Prusti}, {de Bruijne}, {Bailer-Jones}, {Evans}, {Eyer},
  {Jansen}, {Klioner}, {Lammers}, {Luri}, {Mignard}, {Panem}, {Pourbaix},
  {Randich}, {Sartoretti}, {Siddiqui}, {Soubiran}, {Walton}, {Cropper},
  {Drimmel}, {Katz}, {Lattanzi}, {Bakker}, {Cacciari}, {Casta{\~n}eda},
  {Chaoul}, {Cheek}, {De Angeli}, {Fabricius}, {Guerra}, {Holl}, {Masana},
  {Messineo}, {Mowlavi}, {Nienartowicz}, {Panuzzo}, {Portell}, {Riello},
  {Seabroke}, {Tanga}, {Th{\'e}venin}, {Gracia-Abril}, {Comoretto},
  {Garcia-Reinaldos}, {Teyssier}, {Altmann}, {Andrae}, {Audard},
  {Bellas-Velidis}, {Benson}, {Berthier}, {Blomme}, {Burgess}, {Busso},
  {Carry}, {Cellino}, {Clementini}, {Clotet}, {Creevey}, {Davidson}, {De
  Ridder}, {Delchambre}, {Dell'Oro}, {Ducourant},
  {Fern{\'a}ndez-Hern{\'a}ndez}, {Fouesneau}, {Fr{\'e}mat}, {Galluccio},
  {Garc{\'\i}a-Torres}, {Gonz{\'a}lez-N{\'u}{\~n}ez}, {Gonz{\'a}lez-Vidal},
  {Gosset}, {Guy}, {Halbwachs}, {Hambly}, {Harrison}, {Hern{\'a}ndez},
  {Hestroffer}, {Hodgkin}, {Hutton}, {Jasniewicz}, {Jean-Antoine-Piccolo},
  {Jordan}, {Korn}, {Krone-Martins}, {Lanzafame}, {Lebzelter}, {L{\"o}ffler},
  {Manteiga}, {Marrese}, {Mart{\'\i}n-Fleitas}, {Moitinho}, {Mora}, {Muinonen},
  {Osinde}, {Pauwels}, {Petit}, {Recio-Blanco}, {Richards}, {Rimoldini},
  {Sarro}, {Siopis}, {Smith}, {Sozzetti}, {S{\"u}veges}, {Torra}, {van Reeven},
  {Abbas}, {Abreu Aramburu}, {Accart}, {Aerts}, {Altavilla}, {{\'A}lvarez},
  {Alvarez}, {Alves}, {Anderson}, {Andrei}, {Anglada Varela}, {Antiche},
  {Arcay}, {Astraatmadja}, {Bach}, {Baker}, {Balaguer-N{\'u}{\~n}ez}, {Balm},
  {Barache}, {Barata}, {Barbato}, {Barblan}, {Barklem}, {Barrado}, {Barros},
  {Barstow}, {Bartholom{\'e} Mu{\~n}oz}, {Bassilana}, {Becciani}, {Bellazzini},
  {Berihuete}, {Bertone}, {Bianchi}, {Bienaym{\'e}}, {Blanco-Cuaresma}, {Boch},
  {Boeche}, {Bombrun}, {Borrachero}, {Bossini}, {Bouquillon}, {Bourda},
  {Bragaglia}, {Bramante}, {Bressan}, {Brouillet}, {Br{\"u}semeister},
  {Brugaletta}, {Bucciarelli}, {Burlacu}, {Busonero}, {Butkevich}, {Buzzi},
  {Caffau}, {Cancelliere}, {Cannizzaro}, {Cantat-Gaudin}, {Carballo},
  {Carlucci}, {Carrasco}, {Casamiquela}, {Castellani}, {Castro-Ginard},
  {Charlot}, {Chemin}, {Chiavassa}, {Cocozza}, {Costigan}, {Cowell}, {Crifo},
  {Crosta}, {Crowley}, {Cuypers}, {Dafonte}, {Damerdji}, {Dapergolas}, {David},
  {David}, {de Laverny}, {De Luise}, {De March}, {de Martino}, {de Souza}, {de
  Torres}, {Debosscher}, {del Pozo}, {Delbo}, {Delgado}, {Delgado}, {Di
  Matteo}, {Diakite}, {Diener}, {Distefano}, {Dolding}, {Drazinos},
  {Dur{\'a}n}, {Edvardsson}, {Enke}, {Eriksson}, {Esquej}, {Eynard Bontemps},
  {Fabre}, {Fabrizio}, {Faigler}, {Falc{\~a}o}, {Farr{\`a}s Casas}, {Federici},
  {Fedorets}, {Fernique}, {Figueras}, {Filippi}, {Findeisen}, {Fonti},
  {Fraile}, {Fraser}, {Fr{\'e}zouls}, {Gai}, {Galleti}, {Garabato},
  {Garc{\'\i}a-Sedano}, {Garofalo}, {Garralda}, {Gavel}, {Gavras}, {Gerssen},
  {Geyer}, {Giacobbe}, {Gilmore}, {Girona}, {Giuffrida}, {Glass}, {Gomes},
  {Granvik}, {Gueguen}, {Guerrier}, {Guiraud}, {Guti{\'e}rrez-S{\'a}nchez},
  {Hofmann}, {Holland}, {Huckle}, {Hypki}, {Icardi}, {Jan{\ss}en}, {Jevardat de
  Fombelle}, {Jonker}, {Juh{\'a}sz}, {Julbe}, {Karampelas}, {Kewley}, {Klar},
  {Kochoska}, {Kohley}, {Kolenberg}, {Kontizas}, {Kontizas}, {Koposov},
  {Kordopatis}, {Kostrzewa-Rutkowska}, {Koubsky}, {Lambert}, {Lanza}, {Lasne},
  {Lavigne}, {Le Fustec}, {Le Poncin-Lafitte}, {Lebreton}, {Leccia}, {Leclerc},
  {Lecoeur-Taibi}, {Lenhardt}, {Leroux}, {Liao}, {Licata}, {Lindstr{\o}m},
  {Lister}, {Livanou}, {Lobel}, {L{\'o}pez}, {Managau}, {Mann}, {Mantelet},
  {Marchal}, {Marchant}, {Marconi}, {Marinoni}, {Marschalk{\'o}}, {Marshall},
  {Martino}, {Marton}, {Mary}, {Matijevi{\v{c}}}, {Mazeh}, {Messina},
  {Michalik}, {Millar}, {Molina}, {Molinaro}, {Moln{\'a}r}, {Montegriffo},
  {Mor}, {Morbidelli}, {Morel}, {Morris}, {Mulone}, {Muraveva}, {Musella},
  {Nelemans}, {Nicastro}, {Noval}, {O'Mullane}, {Ord{\'e}novic},
  {Ord{\'o}{\~n}ez-Blanco}, {Osborne}, {Pagani}, {Pagano}, {Pailler},
  {Palacin}, {Palaversa}, {Panahi}, {Pawlak}, {Piersimoni}, {Pineau}, {Plachy},
  {Plum}, {Poggio}, {Poujoulet}, {Pr{\v{s}}a}, {Pulone}, {Racero}, {Ragaini},
  {Rambaux}, {Ramos-Lerate}, {Regibo}, {Riclet}, {Ripepi}, {Riva}, {Rivard},
  {Rixon}, {Roegiers}, {Roelens}, {Romero-G{\'o}mez}, {Rowell}, {Royer},
  {Ruiz-Dern}, {Sadowski}, {Sagrist{\`a} Sell{\'e}s}, {Sahlmann}, {Salgado},
  {Salguero}, {Sanna}, {Santana-Ros}, {Sarasso}, {Savietto}, {Schultheis},
  {Sciacca}, {Segol}, {Segovia}, {S{\'e}gransan}, {Shih}, {Siltala}, {Silva},
  {Smart}, {Smith}, {Solano}, {Solitro}, {Sordo}, {Soria Nieto}, {Souchay},
  {Spagna}, {Spoto}, {Stampa}, {Steele}, {Steidelm{\"u}ller}, {Stephenson},
  {Stoev}, {Suess}, {Surdej}, {Szabados}, {Szegedi-Elek}, {Tapiador}, {Taris},
  {Tauran}, {Taylor}, {Teixeira}, {Terrett}, {Teyssand ier}, {Thuillot},
  {Titarenko}, {Torra Clotet}, {Turon}, {Ulla}, {Utrilla}, {Uzzi}, {Vaillant},
  {Valentini}, {Valette}, {van Elteren}, {Van Hemelryck}, {van Leeuwen},
  {Vaschetto}, {Vecchiato}, {Viala}, {Vicente}, {Vogt}, {von Essen}, {Voss},
  {Votruba}, {Voutsinas}, {Walmsley}, {Weiler}, {Wertz}, {Wevems},
  {Wyrzykowski}, {Yoldas}, {{\v{Z}}erjal}, {Ziaeepour}, {Zorec}, {Zschocke},
  {Zucker}, {Zurbach}, \& {Zwitter}}]{Gaia_2018b}
{Gaia Collaboration}, {Helmi}, A., {van Leeuwen}, F., {et~al.}
  2018{\natexlab{a}}, \aap, 616, A12

\bibitem[{{Gaia Collaboration} {et~al.}(2018{\natexlab{b}}){Gaia
  Collaboration}, {Brown}, {Vallenari}, {Prusti}, {de Bruijne}, {Babusiaux},
  {Bailer-Jones}, {Biermann}, {Evans}, {Eyer}, {Jansen}, {Jordi}, {Klioner},
  {Lammers}, {Lindegren}, {Luri}, {Mignard}, {Panem}, {Pourbaix}, {Randich},
  {Sartoretti}, {Siddiqui}, {Soubiran}, {van Leeuwen}, {Walton}, {Arenou},
  {Bastian}, {Cropper}, {Drimmel}, {Katz}, {Lattanzi}, {Bakker}, {Cacciari},
  {Casta{\~n}eda}, {Chaoul}, {Cheek}, {De Angeli}, {Fabricius}, {Guerra},
  {Holl}, {Masana}, {Messineo}, {Mowlavi}, {Nienartowicz}, {Panuzzo},
  {Portell}, {Riello}, {Seabroke}, {Tanga}, {Th{\'e}venin}, {Gracia-Abril},
  {Comoretto}, {Garcia-Reinaldos}, {Teyssier}, {Altmann}, {Andrae}, {Audard},
  {Bellas-Velidis}, {Benson}, {Berthier}, {Blomme}, {Burgess}, {Busso},
  {Carry}, {Cellino}, {Clementini}, {Clotet}, {Creevey}, {Davidson}, {De
  Ridder}, {Delchambre}, {Dell'Oro}, {Ducourant},
  {Fern{\'a}ndez-Hern{\'a}ndez}, {Fouesneau}, {Fr{\'e}mat}, {Galluccio},
  {Garc{\'\i}a-Torres}, {Gonz{\'a}lez-N{\'u}{\~n}ez}, {Gonz{\'a}lez-Vidal},
  {Gosset}, {Guy}, {Halbwachs}, {Hambly}, {Harrison}, {Hern{\'a}ndez},
  {Hestroffer}, {Hodgkin}, {Hutton}, {Jasniewicz}, {Jean-Antoine-Piccolo},
  {Jordan}, {Korn}, {Krone-Martins}, {Lanzafame}, {Lebzelter}, {L{\"o}ffler},
  {Manteiga}, {Marrese}, {Mart{\'\i}n-Fleitas}, {Moitinho}, {Mora}, {Muinonen},
  {Osinde}, {Pancino}, {Pauwels}, {Petit}, {Recio-Blanco}, {Richards},
  {Rimoldini}, {Robin}, {Sarro}, {Siopis}, {Smith}, {Sozzetti}, {S{\"u}veges},
  {Torra}, {van Reeven}, {Abbas}, {Abreu Aramburu}, {Accart}, {Aerts},
  {Altavilla}, {{\'A}lvarez}, {Alvarez}, {Alves}, {Anderson}, {Andrei},
  {Anglada Varela}, {Antiche}, {Antoja}, {Arcay}, {Astraatmadja}, {Bach},
  {Baker}, {Balaguer-N{\'u}{\~n}ez}, {Balm}, {Barache}, {Barata}, {Barbato},
  {Barblan}, {Barklem}, {Barrado}, {Barros}, {Barstow}, {Bartholom{\'e}
  Mu{\~n}oz}, {Bassilana}, {Becciani}, {Bellazzini}, {Berihuete}, {Bertone},
  {Bianchi}, {Bienaym{\'e}}, {Blanco-Cuaresma}, {Boch}, {Boeche}, {Bombrun},
  {Borrachero}, {Bossini}, {Bouquillon}, {Bourda}, {Bragaglia}, {Bramante},
  {Breddels}, {Bressan}, {Brouillet}, {Br{\"u}semeister}, {Brugaletta},
  {Bucciarelli}, {Burlacu}, {Busonero}, {Butkevich}, {Buzzi}, {Caffau},
  {Cancelliere}, {Cannizzaro}, {Cantat-Gaudin}, {Carballo}, {Carlucci},
  {Carrasco}, {Casamiquela}, {Castellani}, {Castro-Ginard}, {Charlot},
  {Chemin}, {Chiavassa}, {Cocozza}, {Costigan}, {Cowell}, {Crifo}, {Crosta},
  {Crowley}, {Cuypers}, {Dafonte}, {Damerdji}, {Dapergolas}, {David}, {David},
  {de Laverny}, {De Luise}, {De March}, {de Martino}, {de Souza}, {de Torres},
  {Debosscher}, {del Pozo}, {Delbo}, {Delgado}, {Delgado}, {Di Matteo},
  {Diakite}, {Diener}, {Distefano}, {Dolding}, {Drazinos}, {Dur{\'a}n},
  {Edvardsson}, {Enke}, {Eriksson}, {Esquej}, {Eynard Bontemps}, {Fabre},
  {Fabrizio}, {Faigler}, {Falc{\~a}o}, {Farr{\`a}s Casas}, {Federici},
  {Fedorets}, {Fernique}, {Figueras}, {Filippi}, {Findeisen}, {Fonti},
  {Fraile}, {Fraser}, {Fr{\'e}zouls}, {Gai}, {Galleti}, {Garabato},
  {Garc{\'\i}a-Sedano}, {Garofalo}, {Garralda}, {Gavel}, {Gavras}, {Gerssen},
  {Geyer}, {Giacobbe}, {Gilmore}, {Girona}, {Giuffrida}, {Glass}, {Gomes},
  {Granvik}, {Gueguen}, {Guerrier}, {Guiraud}, {Guti{\'e}rrez-S{\'a}nchez},
  {Haigron}, {Hatzidimitriou}, {Hauser}, {Haywood}, {Heiter}, {Helmi}, {Heu},
  {Hilger}, {Hobbs}, {Hofmann}, {Holland}, {Huckle}, {Hypki}, {Icardi},
  {Jan{\ss}en}, {Jevardat de Fombelle}, {Jonker}, {Juh{\'a}sz}, {Julbe},
  {Karampelas}, {Kewley}, {Klar}, {Kochoska}, {Kohley}, {Kolenberg},
  {Kontizas}, {Kontizas}, {Koposov}, {Kordopatis}, {Kostrzewa-Rutkowska},
  {Koubsky}, {Lambert}, {Lanza}, {Lasne}, {Lavigne}, {Le Fustec}, {Le
  Poncin-Lafitte}, {Lebreton}, {Leccia}, {Leclerc}, {Lecoeur-Taibi},
  {Lenhardt}, {Leroux}, {Liao}, {Licata}, {Lindstr{\o}m}, {Lister}, {Livanou},
  {Lobel}, {L{\'o}pez}, {Managau}, {Mann}, {Mantelet}, {Marchal}, {Marchant},
  {Marconi}, {Marinoni}, {Marschalk{\'o}}, {Marshall}, {Martino}, {Marton},
  {Mary}, {Massari}, {Matijevi{\v{c}}}, {Mazeh}, {McMillan}, {Messina},
  {Michalik}, {Millar}, {Molina}, {Molinaro}, {Moln{\'a}r}, {Montegriffo},
  {Mor}, {Morbidelli}, {Morel}, {Morris}, {Mulone}, {Muraveva}, {Musella},
  {Nelemans}, {Nicastro}, {Noval}, {O'Mullane}, {Ord{\'e}novic},
  {Ord{\'o}{\~n}ez-Blanco}, {Osborne}, {Pagani}, {Pagano}, {Pailler},
  {Palacin}, {Palaversa}, {Panahi}, {Pawlak}, {Piersimoni}, {Pineau}, {Plachy},
  {Plum}, {Poggio}, {Poujoulet}, {Pr{\v{s}}a}, {Pulone}, {Racero}, {Ragaini},
  {Rambaux}, {Ramos-Lerate}, {Regibo}, {Reyl{\'e}}, {Riclet}, {Ripepi}, {Riva},
  {Rivard}, {Rixon}, {Roegiers}, {Roelens}, {Romero-G{\'o}mez}, {Rowell},
  {Royer}, {Ruiz-Dern}, {Sadowski}, {Sagrist{\`a} Sell{\'e}s}, {Sahlmann},
  {Salgado}, {Salguero}, {Sanna}, {Santana-Ros}, {Sarasso}, {Savietto},
  {Schultheis}, {Sciacca}, {Segol}, {Segovia}, {S{\'e}gransan}, {Shih},
  {Siltala}, {Silva}, {Smart}, {Smith}, {Solano}, {Solitro}, {Sordo}, {Soria
  Nieto}, {Souchay}, {Spagna}, {Spoto}, {Stampa}, {Steele},
  {Steidelm{\"u}ller}, {Stephenson}, {Stoev}, {Suess}, {Surdej}, {Szabados},
  {Szegedi-Elek}, {Tapiador}, {Taris}, {Tauran}, {Taylor}, {Teixeira},
  {Terrett}, {Teyssand ier}, {Thuillot}, {Titarenko}, {Torra Clotet}, {Turon},
  {Ulla}, {Utrilla}, {Uzzi}, {Vaillant}, {Valentini}, {Valette}, {van Elteren},
  {Van Hemelryck}, {van Leeuwen}, {Vaschetto}, {Vecchiato}, {Veljanoski},
  {Viala}, {Vicente}, {Vogt}, {von Essen}, {Voss}, {Votruba}, {Voutsinas},
  {Walmsley}, {Weiler}, {Wertz}, {Wevers}, {Wyrzykowski}, {Yoldas},
  {{\v{Z}}erjal}, {Ziaeepour}, {Zorec}, {Zschocke}, {Zucker}, {Zurbach}, \&
  {Zwitter}}]{Gaia_2018a}
{Gaia Collaboration}, {Brown}, A.~G.~A., {Vallenari}, A., {et~al.}
  2018{\natexlab{b}}, \aap, 616, A1

\bibitem[{{Girardi} {et~al.}(2000){Girardi}, {Bressan}, {Bertelli}, \&
  {Chiosi}}]{Girardi_2000}
{Girardi}, L., {Bressan}, A., {Bertelli}, G., \& {Chiosi}, C. 2000, \aaps, 141,
  371

\bibitem[{{Gravity Collaboration} {et~al.}(2018){Gravity Collaboration},
  {Abuter}, {Amorim}, {Anugu}, {Baub{\"o}ck}, {Benisty}, {Berger}, {Blind},
  {Bonnet}, {Brandner}, {Buron}, {Collin}, {Chapron}, {Cl{\'e}net}, {Coud{\'e}
  Du Foresto}, {de Zeeuw}, {Deen}, {Delplancke-Str{\"o}bele}, {Dembet},
  {Dexter}, {Duvert}, {Eckart}, {Eisenhauer}, {Finger}, {F{\"o}rster
  Schreiber}, {F{\'e}dou}, {Garcia}, {Garcia Lopez}, {Gao}, {Gendron},
  {Genzel}, {Gillessen}, {Gordo}, {Habibi}, {Haubois}, {Haug}, {Hau{\ss}mann},
  {Henning}, {Hippler}, {Horrobin}, {Hubert}, {Hubin}, {Jimenez Rosales},
  {Jochum}, {Jocou}, {Kaufer}, {Kellner}, {Kendrew}, {Kervella}, {Kok},
  {Kulas}, {Lacour}, {Lapeyr{\`e}re}, {Lazareff}, {Le Bouquin}, {L{\'e}na},
  {Lippa}, {Lenzen}, {M{\'e}rand}, {M{\"u}ler}, {Neumann}, {Ott}, {Palanca},
  {Paumard}, {Pasquini}, {Perraut}, {Perrin}, {Pfuhl}, {Plewa}, {Rabien},
  {Ram{\'\i}rez}, {Ramos}, {Rau}, {Rodr{\'\i}guez-Coira}, {Rohloff}, {Rousset},
  {Sanchez-Bermudez}, {Scheithauer}, {Sch{\"o}ller}, {Schuler}, {Spyromilio},
  {Straub}, {Straubmeier}, {Sturm}, {Tacconi}, {Tristram}, {Vincent}, {von
  Fellenberg}, {Wank}, {Waisberg}, {Widmann}, {Wieprecht}, {Wiest},
  {Wiezorrek}, {Woillez}, {Yazici}, {Ziegler}, \& {Zins}}]{Gravity_2018}
{Gravity Collaboration}, {Abuter}, R., {Amorim}, A., {et~al.} 2018, \aap, 615,
  L15

\bibitem[{{Griffen} {et~al.}(2010){Griffen}, {Drinkwater}, {Thomas}, {Helly},
  \& {Pimbblet}}]{Griffen_2010}
{Griffen}, B.~F., {Drinkwater}, M.~J., {Thomas}, P.~A., {Helly}, J.~C., \&
  {Pimbblet}, K.~A. 2010, \mnras, 405, 375

\bibitem[{{Hambly} {et~al.}(2018){Hambly}, {Cropper}, {Boudreault}, {Crowley},
  {Kohley}, {de Bruijne}, {Dolding}, {Fabricius}, {Seabroke}, {Davidson},
  {Rowell}, {Collins}, {Cross}, {Mart{\'\i}n-Fleitas}, {Baker}, {Smith},
  {Sartoretti}, {Marchal}, {Katz}, {De Angeli}, {Busso}, {Riello}, {Allende
  Prieto}, {Els}, {Corcione}, {Masana}, {Luri}, {Chassat}, {Fusero},
  {Pasquier}, {V{\'e}tel}, {Sarri}, \& {Gare}}]{Hambly_2018}
{Hambly}, N.~C., {Cropper}, M., {Boudreault}, S., {et~al.} 2018, \aap, 616, A15

\bibitem[{{Harris}(1996)}]{Harris_2010}
{Harris}, W.~E. 1996, \aj, 112, 1487

\bibitem[{{Hasselquist} {et~al.}(2017){Hasselquist}, {Shetrone}, {Smith},
  {Holtzman}, {McWilliam}, {Fern{\'a}ndez-Trincado}, {Beers}, {Majewski},
  {Nidever}, {Tang}, {Tissera}, {Fern{\'a}ndez Alvar}, {Allende Prieto},
  {Almeida}, {Anguiano}, {Battaglia}, {Carigi}, {Delgado Inglada},
  {Frinchaboy}, {Garc{\'\i}a-Hern{\'a}ndez}, {Geisler}, {Minniti}, {Placco},
  {Schultheis}, {Sobeck}, \& {Villanova}}]{Hasselquist_2017}
{Hasselquist}, S., {Shetrone}, M., {Smith}, V., {et~al.} 2017, \apj, 845, 162

\bibitem[{{Hughes} {et~al.}(2019){Hughes}, {Pfeffer}, {Martig}, {Bastian},
  {Crain}, {Kruijssen}, \& {Reina-Campos}}]{Hughes_2019}
{Hughes}, M.~E., {Pfeffer}, J., {Martig}, M., {et~al.} 2019, \mnras, 482, 2795

\bibitem[{{Juri{\'c}} {et~al.}(2008){Juri{\'c}}, {Ivezi{\'c}}, {Brooks},
  {Lupton}, {Schlegel}, {Finkbeiner}, {Padmanabhan}, {Bond}, {Sesar},
  {Rockosi}, {Knapp}, {Gunn}, {Sumi}, {Schneider}, {Barentine}, {Brewington},
  {Brinkmann}, {Fukugita}, {Harvanek}, {Kleinman}, {Krzesinski}, {Long},
  {Neilsen}, {Nitta}, {Snedden}, \& {York}}]{Juric_2008}
{Juri{\'c}}, M., {Ivezi{\'c}}, {\v{Z}}., {Brooks}, A., {et~al.} 2008, \apj,
  673, 864

\bibitem[{{Keller} {et~al.}(2020){Keller}, {Kruijssen}, {Pfeffer},
  {Reina-Campos}, {Bastian}, {Trujillo-Gomez}, {Hughes}, \&
  {Crain}}]{Keller_2020}
{Keller}, B.~W., {Kruijssen}, J.~M.~D., {Pfeffer}, J., {et~al.} 2020, \mnras,
  495, 4248

\bibitem[{{Kim} {et~al.}(2018){Kim}, {Ma}, {Grudi{\'c}}, {Hopkins}, {Hayward},
  {Wetzel}, {Faucher-Gigu{\`e}re}, {Kere{\v{s}}}, {Garrison-Kimmel}, \&
  {Murray}}]{Kim_2018}
{Kim}, J.-h., {Ma}, X., {Grudi{\'c}}, M.~Y., {et~al.} 2018, \mnras, 474, 4232

\bibitem[{{Kirby} {et~al.}(2013){Kirby}, {Cohen}, {Guhathakurta}, {Cheng},
  {Bullock}, \& {Gallazzi}}]{Kirby_2013}
{Kirby}, E.~N., {Cohen}, J.~G., {Guhathakurta}, P., {et~al.} 2013, \apj, 779,
  102

\bibitem[{{Kravtsov} \& {Gnedin}(2005)}]{Kravtsov_2005}
{Kravtsov}, A.~V., \& {Gnedin}, O.~Y. 2005, \apj, 623, 650

\bibitem[{{Kruijssen}(2015)}]{Kruijssen_2015}
{Kruijssen}, J.~M.~D. 2015, \mnras, 454, 1658

\bibitem[{{Kruijssen}(2019)}]{Kruijssen_2019a}
---. 2019, \mnras, 486, L20

\bibitem[{{Kruijssen} {et~al.}(2019{\natexlab{a}}){Kruijssen}, {Pfeffer},
  {Crain}, \& {Bastian}}]{Kruijssen_2019b}
{Kruijssen}, J.~M.~D., {Pfeffer}, J.~L., {Crain}, R.~A., \& {Bastian}, N.
  2019{\natexlab{a}}, \mnras, 486, 3134

\bibitem[{{Kruijssen} {et~al.}(2019{\natexlab{b}}){Kruijssen}, {Pfeffer},
  {Reina-Campos}, {Crain}, \& {Bastian}}]{Kruijssen_2019c}
{Kruijssen}, J.~M.~D., {Pfeffer}, J.~L., {Reina-Campos}, M., {Crain}, R.~A., \&
  {Bastian}, N. 2019{\natexlab{b}}, \mnras, 486, 3180

\bibitem[{{Kruijssen} {et~al.}(2020){Kruijssen}, {Pfeffer}, {Chevance},
  {Bonaca}, {Trujillo-Gomez}, {Bastian}, {Reina-Campos}, {Crain}, \&
  {Hughes}}]{Kruijssen_2020}
{Kruijssen}, J.~M.~D., {Pfeffer}, J.~L., {Chevance}, M., {et~al.} 2020, \mnras,
  498, 2472

\bibitem[{{Larsen} {et~al.}(2012){Larsen}, {Brodie}, \&
  {Strader}}]{Larsen_2012}
{Larsen}, S.~S., {Brodie}, J.~P., \& {Strader}, J. 2012, \aap, 546, A53

\bibitem[{{Law} \& {Majewski}(2010)}]{Law_2010}
{Law}, D.~R., \& {Majewski}, S.~R. 2010, \apj, 714, 229

\bibitem[{{Leaman} {et~al.}(2013){Leaman}, {VandenBerg}, \&
  {Mendel}}]{Leaman_2013}
{Leaman}, R., {VandenBerg}, D.~A., \& {Mendel}, J.~T. 2013, \mnras, 436, 122

\bibitem[{{Letarte} {et~al.}(2006){Letarte}, {Hill}, {Jablonka}, {Tolstoy},
  {Fran{\c{c}}ois}, \& {Meylan}}]{Letarte_2006}
{Letarte}, B., {Hill}, V., {Jablonka}, P., {et~al.} 2006, \aap, 453, 547

\bibitem[{{Li} \& {Gnedin}(2014)}]{Li_2014}
{Li}, H., \& {Gnedin}, O.~Y. 2014, \apj, 796, 10

\bibitem[{{Li} \& {Gnedin}(2019)}]{Li_2019}
---. 2019, \mnras, 486, 4030

\bibitem[{{Lindegren} {et~al.}(2018){Lindegren}, {Hern{\'a}ndez}, {Bombrun},
  {Klioner}, {Bastian}, {Ramos-Lerate}, {de Torres}, {Steidelm{\"u}ller},
  {Stephenson}, {Hobbs}, {Lammers}, {Biermann}, {Geyer}, {Hilger}, {Michalik},
  {Stampa}, {McMillan}, {Casta{\~n}eda}, {Clotet}, {Comoretto}, {Davidson},
  {Fabricius}, {Gracia}, {Hambly}, {Hutton}, {Mora}, {Portell}, {van Leeuwen},
  {Abbas}, {Abreu}, {Altmann}, {Andrei}, {Anglada}, {Balaguer-N{\'u}{\~n}ez},
  {Barache}, {Becciani}, {Bertone}, {Bianchi}, {Bouquillon}, {Bourda},
  {Br{\"u}semeister}, {Bucciarelli}, {Busonero}, {Buzzi}, {Cancelliere},
  {Carlucci}, {Charlot}, {Cheek}, {Crosta}, {Crowley}, {de Bruijne}, {de
  Felice}, {Drimmel}, {Esquej}, {Fienga}, {Fraile}, {Gai}, {Garralda},
  {Gonz{\'a}lez-Vidal}, {Guerra}, {Hauser}, {Hofmann}, {Holl}, {Jordan},
  {Lattanzi}, {Lenhardt}, {Liao}, {Licata}, {Lister}, {L{\"o}ffler},
  {Marchant}, {Martin-Fleitas}, {Messineo}, {Mignard}, {Morbidelli}, {Poggio},
  {Riva}, {Rowell}, {Salguero}, {Sarasso}, {Sciacca}, {Siddiqui}, {Smart},
  {Spagna}, {Steele}, {Taris}, {Torra}, {van Elteren}, {van Reeven}, \&
  {Vecchiato}}]{Lindegren_2018}
{Lindegren}, L., {Hern{\'a}ndez}, J., {Bombrun}, A., {et~al.} 2018, \aap, 616,
  A2

\bibitem[{{Luri} {et~al.}(2018){Luri}, {Brown}, {Sarro}, {Arenou},
  {Bailer-Jones}, {Castro-Ginard}, {de Bruijne}, {Prusti}, {Babusiaux}, \&
  {Delgado}}]{Luri_2018}
{Luri}, X., {Brown}, A.~G.~A., {Sarro}, L.~M., {et~al.} 2018, \aap, 616, A9

\bibitem[{{Mackey} \& {Gilmore}(2004)}]{Mackey_2004}
{Mackey}, A.~D., \& {Gilmore}, G.~F. 2004, \mnras, 355, 504

\bibitem[{{Mar{\'\i}n-Franch} {et~al.}(2009){Mar{\'\i}n-Franch}, {Aparicio},
  {Piotto}, {Rosenberg}, {Chaboyer}, {Sarajedini}, {Siegel}, {Anderson},
  {Bedin}, {Dotter}, {Hempel}, {King}, {Majewski}, {Milone}, {Paust}, \&
  {Reid}}]{Marin_2009}
{Mar{\'\i}n-Franch}, A., {Aparicio}, A., {Piotto}, G., {et~al.} 2009, \apj,
  694, 1498

\bibitem[{{Massari} {et~al.}(2019){Massari}, {Koppelman}, \&
  {Helmi}}]{Massari_2019}
{Massari}, D., {Koppelman}, H.~H., \& {Helmi}, A. 2019, \aap, 630, L4

\bibitem[{McKinney(2010)}]{McKinney_2010}
McKinney, W. 2010, in Proceedings of the 9th Python in Science Conference
  ({SciPy})

\bibitem[{{McMillan}(2017)}]{McMillan_2017}
{McMillan}, P.~J. 2017, \mnras, 465, 76

\bibitem[{{Miyamoto} \& {Nagai}(1975)}]{Miyamoto_1975}
{Miyamoto}, M., \& {Nagai}, R. 1975, \pasj, 27, 533

\bibitem[{{Mo} {et~al.}(2010){Mo}, {van den Bosch}, \& {White}}]{Mo_2010}
{Mo}, H., {van den Bosch}, F.~C., \& {White}, S. 2010, {Galaxy Formation and
  Evolution}

\bibitem[{{Mottini} {et~al.}(2008){Mottini}, {Wallerstein}, \&
  {McWilliam}}]{Mottini_2008}
{Mottini}, M., {Wallerstein}, G., \& {McWilliam}, A. 2008, \aj, 136, 614

\bibitem[{{Muratov} \& {Gnedin}(2010)}]{Muratov_2010}
{Muratov}, A.~L., \& {Gnedin}, O.~Y. 2010, \apj, 718, 1266

\bibitem[{{Myeong} {et~al.}(2018){Myeong}, {Evans}, {Belokurov}, {Sanders}, \&
  {Koposov}}]{Myeong_2018}
{Myeong}, G.~C., {Evans}, N.~W., {Belokurov}, V., {Sanders}, J.~L., \&
  {Koposov}, S.~E. 2018, \apjl, 863, L28

\bibitem[{{Navarro} {et~al.}(1996){Navarro}, {Frenk}, \&
  {White}}]{Navarro_1996}
{Navarro}, J.~F., {Frenk}, C.~S., \& {White}, S. D.~M. 1996, \apj, 462, 563

\bibitem[{{Ochsenbein} {et~al.}(2000){Ochsenbein}, {Bauer}, \&
  {Marcout}}]{Ochsenbein_2000}
{Ochsenbein}, F., {Bauer}, P., \& {Marcout}, J. 2000, \aaps, 143, 23

\bibitem[{{Pascale} {et~al.}(2018){Pascale}, {Posti}, {Nipoti}, \&
  {Binney}}]{Pascale_2018}
{Pascale}, R., {Posti}, L., {Nipoti}, C., \& {Binney}, J. 2018, \mnras, 480,
  927

\bibitem[{{Patel} {et~al.}(2013{\natexlab{a}}){Patel}, {van Dokkum}, {Franx},
  {Quadri}, {Muzzin}, {Marchesini}, {Williams}, {Holden}, \&
  {Stefanon}}]{Patel_2013a}
{Patel}, S.~G., {van Dokkum}, P.~G., {Franx}, M., {et~al.} 2013{\natexlab{a}},
  \apj, 766, 15

\bibitem[{{Patel} {et~al.}(2013{\natexlab{b}}){Patel}, {Fumagalli}, {Franx},
  {van Dokkum}, {van der Wel}, {Leja}, {Labb{\'e}}, {Brammer}, {Skelton},
  {Momcheva}, {Whitaker}, {Lundgren}, {Muzzin}, {Quadri}, {Nelson}, {Wake}, \&
  {Rix}}]{Patel_2013b}
{Patel}, S.~G., {Fumagalli}, M., {Franx}, M., {et~al.} 2013{\natexlab{b}},
  \apj, 778, 115

\bibitem[{{Pfeffer} {et~al.}(2018){Pfeffer}, {Kruijssen}, {Crain}, \&
  {Bastian}}]{Pfeffer_2018}
{Pfeffer}, J., {Kruijssen}, J.~M.~D., {Crain}, R.~A., \& {Bastian}, N. 2018,
  \mnras, 475, 4309

\bibitem[{{Pfeffer} {et~al.}(2020){Pfeffer}, {Trujillo-Gomez}, {Kruijssen},
  {Crain}, {Hughes}, {Reina-Campos}, \& {Bastian}}]{Pfeffer_2020}
{Pfeffer}, J.~L., {Trujillo-Gomez}, S., {Kruijssen}, J.~M.~D., {et~al.} 2020,
  \mnras, 499, 4863

\bibitem[{{Pietrinferni} {et~al.}(2004){Pietrinferni}, {Cassisi}, {Salaris}, \&
  {Castelli}}]{Pietrinferni_2004}
{Pietrinferni}, A., {Cassisi}, S., {Salaris}, M., \& {Castelli}, F. 2004, \apj,
  612, 168

\bibitem[{{Renaud} {et~al.}(2017){Renaud}, {Agertz}, \& {Gieles}}]{Renaud_2017}
{Renaud}, F., {Agertz}, O., \& {Gieles}, M. 2017, \mnras, 465, 3622

\bibitem[{{Rutledge} {et~al.}(1997{\natexlab{a}}){Rutledge}, {Hesser}, \&
  {Stetson}}]{Rutledge_1997b}
{Rutledge}, G.~A., {Hesser}, J.~E., \& {Stetson}, P.~B. 1997{\natexlab{a}},
  \pasp, 109, 907

\bibitem[{{Rutledge} {et~al.}(1997{\natexlab{b}}){Rutledge}, {Hesser},
  {Stetson}, {Mateo}, {Simard}, {Bolte}, {Friel}, \& {Copin}}]{Rutledge_1997a}
{Rutledge}, G.~A., {Hesser}, J.~E., {Stetson}, P.~B., {et~al.}
  1997{\natexlab{b}}, \pasp, 109, 883

\bibitem[{{Sbordone} {et~al.}(2007){Sbordone}, {Bonifacio}, {Buonanno},
  {Marconi}, {Monaco}, \& {Zaggia}}]{Sbordone_2007}
{Sbordone}, L., {Bonifacio}, P., {Buonanno}, R., {et~al.} 2007, \aap, 465, 815

\bibitem[{{Sch{\"o}nrich} {et~al.}(2010){Sch{\"o}nrich}, {Binney}, \&
  {Dehnen}}]{Schonrich_2010}
{Sch{\"o}nrich}, R., {Binney}, J., \& {Dehnen}, W. 2010, \mnras, 403, 1829

\bibitem[{Seabold \& Perktold(2010)}]{Seabold_2010}
Seabold, S., \& Perktold, J. 2010, in 9th Python in Science Conference

\bibitem[{{Searle} \& {Zinn}(1978)}]{Searle_1978}
{Searle}, L., \& {Zinn}, R. 1978, \apj, 225, 357

\bibitem[{{Simon}(2019)}]{Simon_2019}
{Simon}, J.~D. 2019, \araa, 57, 375

\bibitem[{{Sohn} {et~al.}(2018){Sohn}, {Watkins}, {Fardal}, {van der Marel},
  {Deason}, {Besla}, \& {Bellini}}]{Sohn_2018}
{Sohn}, S.~T., {Watkins}, L.~L., {Fardal}, M.~A., {et~al.} 2018, \apj, 862, 52

\bibitem[{{Trujillo-Gomez} {et~al.}(2021){Trujillo-Gomez}, {Kruijssen},
  {Reina-Campos}, {Pfeffer}, {Keller}, {Crain}, {Bastian}, \&
  {Hughes}}]{Trujillo-Gomez_2021}
{Trujillo-Gomez}, S., {Kruijssen}, J.~M.~D., {Reina-Campos}, M., {et~al.} 2021,
  \mnras, 503, 31

\bibitem[{{van der Walt} {et~al.}(2011){van der Walt}, {Colbert}, \&
  {Varoquaux}}]{vanderWalt_2011}
{van der Walt}, S., {Colbert}, S.~C., \& {Varoquaux}, G. 2011, Computing in
  Science and Engineering, 13, 22

\bibitem[{{VandenBerg} {et~al.}(2013){VandenBerg}, {Brogaard}, {Leaman}, \&
  {Casagrand e}}]{VandenBerg_2013}
{VandenBerg}, D.~A., {Brogaard}, K., {Leaman}, R., \& {Casagrand e}, L. 2013,
  \apj, 775, 134

\bibitem[{{Vasiliev} \& {Belokurov}(2020)}]{Vasiliev_2020}
{Vasiliev}, E., \& {Belokurov}, V. 2020, \mnras, 497, 4162

\bibitem[{{Wagner-Kaiser} {et~al.}(2017){Wagner-Kaiser}, {Mackey},
  {Sarajedini}, {Chaboyer}, {Cohen}, {Yang}, {Cummings}, {Geisler}, \&
  {Grocholski}}]{Wagner-Kaiser_2017}
{Wagner-Kaiser}, R., {Mackey}, D., {Sarajedini}, A., {et~al.} 2017, \mnras,
  471, 3347

\bibitem[{{Wang} {et~al.}(2020){Wang}, {Han}, {Cautun}, {Li}, \&
  {Ishigaki}}]{Wang_2020}
{Wang}, W., {Han}, J., {Cautun}, M., {Li}, Z., \& {Ishigaki}, M.~N. 2020,
  Science China Physics, Mechanics, and Astronomy, 63, 109801

\bibitem[{{Wenger} {et~al.}(2000){Wenger}, {Ochsenbein}, {Egret}, {Dubois},
  {Bonnarel}, {Borde}, {Genova}, {Jasniewicz}, {Lalo{\"e}}, {Lesteven}, \&
  {Monier}}]{Wenger_2000}
{Wenger}, M., {Ochsenbein}, F., {Egret}, D., {et~al.} 2000, \aaps, 143, 9

\bibitem[{{West} {et~al.}(2004){West}, {C{\^o}t{\'e}}, {Marzke}, \&
  {Jord{\'a}n}}]{West_2004}
{West}, M.~J., {C{\^o}t{\'e}}, P., {Marzke}, R.~O., \& {Jord{\'a}n}, A. 2004,
  \nat, 427, 31

\bibitem[{{Wright}(2006)}]{Wright_2006}
{Wright}, E.~L. 2006, \pasp, 118, 1711

\bibitem[{{Zinn} \& {West}(1984)}]{Zinn_1984}
{Zinn}, R., \& {West}, M.~J. 1984, \apjs, 55, 45

\end{thebibliography}
\bibliographystyle{aasjournal}

\begin{longrotatetable}
\begin{deluxetable*}{llDDDDDDDDDD}
\tabletypesize{\scriptsize}
\tablenum{1}
\tablecaption{Input Astrometric, Distance, and Radial Velocity Data}
\tablehead{
\colhead{NGC} &
\colhead{Alternate} &
\twocolhead{$\alpha$} &
\twocolhead{$\delta$} &
\twocolhead{$\mu_{\alpha}$} &
\twocolhead{$\mu_{\delta}$} &
\twocolhead{$\varpi$} &
\twocolhead{$\sigma_{\alpha\delta}$} &
\twocolhead{$\sigma_{\alpha\varpi}$} &
\twocolhead{$\sigma_{\delta\varpi}$} &
\twocolhead{$d$} &
\twocolhead{$v_{r}$} \\
\colhead{Name} &
\colhead{Name} &
\twocolhead{[deg]} &
\twocolhead{[deg]} &
\twocolhead{[mas yr$^{-1}$]} &
\twocolhead{[mas yr$^{-1}$]} &
\twocolhead{[mas]} &
\twocolhead{} &
\twocolhead{} &
\twocolhead{} &
\twocolhead{[kpc]} &
\twocolhead{[km s$^{-1}$]}}
\decimals
\startdata
NGC 104 & 47 Tuc & 6.0194 & -72.0821 & 5.2477^{+0.0016}_{-0.0016}  & -2.5189^{+0.0015}_{-0.0015} & 0.1959^{+0.0002}_{-0.0002} & -0.06 & -0.01 & -0.01 & 4.3^{+0.1}_{-0.1} & -18.7^{+0.2}_{-0.2} \\
NGC 288 & $\ldots$ & 13.1879 & -26.5858 & 4.2385^{+0.0035}_{-0.0035}  & -5.6470^{+0.0026}_{-0.0026} & 0.1401^{+0.0021}_{-0.0021} & 0.25 & 0.15 & -0.13 & 8.1^{+0.2}_{-0.2} & -46.6^{+0.4}_{-0.4} \\
NGC 362 & $\ldots$ & 15.8099 & -70.8489 & 6.6954^{+0.0045}_{-0.0045}  & -2.5184^{+0.0034}_{-0.0034} & 0.0788^{+0.0012}_{-0.0012} & -0.09 & -0.04 & -0.12 & 8.3^{+0.2}_{-0.2} & 223.5^{+0.5}_{-0.5} \\
NGC 1261 & $\ldots$ & 48.067543 & -55.216224 & 1.6900^{+0.0400}_{-0.0400}  & -2.1100^{+0.0400}_{-0.0400} & . $\ldots$ & . $\ldots$ & . $\ldots$ & . $\ldots$ & 16.0^{+0.4}_{-0.4} & 68.2^{+4.6}_{-4.6} \\
NGC 1851 & $\ldots$ & 78.528 & -40.0456 & 2.1308^{+0.0037}_{-0.0037}  & -0.6220^{+0.0040}_{-0.0040} & 0.0298^{+0.0011}_{-0.0011} & -0.09 & 0.06 & -0.07 & 12.2^{+0.3}_{-0.3} & 320.9^{+1.0}_{-1.0} \\
NGC 2298 & $\ldots$ & 102.2464 & -36.0046 & 3.2762^{+0.0060}_{-0.0060}  & -2.1913^{+0.0061}_{-0.0061} & 0.0791^{+0.0019}_{-0.0019} & 0.07 & 0.08 & -0.07 & 10.6^{+0.2}_{-0.2} & 148.9^{+1.2}_{-1.2} \\
NGC 2808 & $\ldots$ & 138.0071 & -64.8645 & 1.0032^{+0.0032}_{-0.0032}  & 0.2785^{+0.0032}_{-0.0032} & 0.0560^{+0.0006}_{-0.0006} & -0.08 & 0.05 & -0.01 & 9.3^{+0.2}_{-0.2} & 93.6^{+2.4}_{-2.4} \\
NGC 3201 & $\ldots$ & 154.3987 & -46.4125 & 8.3344^{+0.0021}_{-0.0021}  & -1.9895^{+0.0020}_{-0.0020} & 0.1724^{+0.0006}_{-0.0006} & 0.12 & 0.04 & -0.02 & 5.1^{+0.1}_{-0.1} & 494.0^{+0.2}_{-0.2} \\
NGC 4147 & $\ldots$ & 182.52626 & 18.542638 & -1.7800^{+0.0400}_{-0.0400}  & -2.1000^{+0.0400}_{-0.0400} & . $\ldots$ & . $\ldots$ & . $\ldots$ & . $\ldots$ & 18.8^{+0.4}_{-0.4} & 183.2^{+0.7}_{-0.7} \\
NGC 4590 & M 68 & 189.8651 & -26.7454 & -2.7640^{+0.0050}_{-0.0050}  & 1.7916^{+0.0039}_{-0.0039} & 0.0664^{+0.0025}_{-0.0025} & -0.29 & -0.0 & 0.13 & 10.1^{+0.2}_{-0.2} & -95.2^{+0.4}_{-0.4} \\
NGC 4833 & $\ldots$ & 194.8978 & -70.8718 & -8.3147^{+0.0036}_{-0.0036}  & -0.9366^{+0.0029}_{-0.0029} & 0.1163^{+0.0010}_{-0.0010} & 0.06 & 0.05 & 0.11 & 5.9^{+0.1}_{-0.1} & 200.2^{+1.2}_{-1.2} \\
NGC 5024 & M 53 & 198.2262 & 18.1661 & -0.1466^{+0.0045}_{-0.0045}  & -1.3514^{+0.0032}_{-0.0032} & 0.0143^{+0.0018}_{-0.0018} & -0.28 & -0.12 & 0.08 & 18.4^{+0.4}_{-0.4} & -79.1^{+4.1}_{-4.1} \\
NGC 5053 & $\ldots$ & 199.1124 & 17.7008 & -0.3591^{+0.0071}_{-0.0071}  & -1.2586^{+0.0048}_{-0.0048} & 0.0064^{+0.0040}_{-0.0040} & -0.32 & 0.13 & -0.17 & 16.2^{+0.4}_{-0.4} & 44.0^{+0.4}_{-0.4} \\
NGC 5139 & $\omega$ Cen & 201.7876 & -47.4515 & -3.1925^{+0.0022}_{-0.0022}  & -6.7445^{+0.0019}_{-0.0019} & 0.1237^{+0.0011}_{-0.0011} & -0.03 & -0.04 & 0.17 & 5.1^{+0.1}_{-0.1} & 232.3^{+0.5}_{-0.5} \\
NGC 5272 & M 3 & 205.5486 & 28.376 & -0.1127^{+0.0029}_{-0.0029}  & -2.6274^{+0.0022}_{-0.0022} & 0.0265^{+0.0010}_{-0.0010} & -0.03 & -0.01 & -0.05 & 10.0^{+0.2}_{-0.2} & -148.6^{+0.4}_{-0.4} \\
NGC 5286 & $\ldots$ & 206.6136 & -51.3723 & 0.1836^{+0.0076}_{-0.0076}  & -0.1477^{+0.0068}_{-0.0068} & 0.0168^{+0.0025}_{-0.0025} & -0.01 & -0.02 & 0.08 & 10.7^{+0.2}_{-0.2} & 58.3^{+3.2}_{-3.2} \\
NGC 5466 & $\ldots$ & 211.3614 & 28.5331 & -5.4044^{+0.0042}_{-0.0042}  & -0.7907^{+0.0041}_{-0.0041} & 0.0210^{+0.0021}_{-0.0021} & 0.07 & 0.04 & 0.15 & 16.6^{+0.4}_{-0.4} & 107.7^{+0.3}_{-0.3} \\
NGC 5904 & M 5 & 229.6394 & 2.0766 & 4.0613^{+0.0032}_{-0.0032}  & -9.8610^{+0.0029}_{-0.0029} & 0.1135^{+0.0010}_{-0.0010} & 0.03 & -0.07 & 0.09 & 7.3^{+0.2}_{-0.2} & 52.1^{+0.5}_{-0.5} \\
NGC 5927 & $\ldots$ & 232.0065 & -50.6694 & -5.0470^{+0.0060}_{-0.0060}  & -3.2325^{+0.0055}_{-0.0055} & 0.0996^{+0.0021}_{-0.0021} & -0.08 & -0.0 & -0.01 & 7.4^{+0.2}_{-0.2} & -115.7^{+3.1}_{-3.1} \\
NGC 5986 & $\ldots$ & 236.5211 & -37.7826 & -4.2217^{+0.0084}_{-0.0084}  & -4.5515^{+0.0065}_{-0.0065} & 0.0718^{+0.0031}_{-0.0031} & -0.18 & -0.1 & -0.01 & 10.3^{+0.2}_{-0.2} & 88.9^{+3.7}_{-3.7} \\
NGC 6093 & M 80 & 244.2564 & -22.9723 & -2.9469^{+0.0090}_{-0.0090}  & -5.5613^{+0.0073}_{-0.0073} & 0.0558^{+0.0030}_{-0.0030} & 0.01 & -0.1 & 0.06 & 8.7^{+0.2}_{-0.2} & 9.3^{+3.1}_{-3.1} \\
NGC 6121 & M 4 & 245.8976 & -26.5279 & -12.4956^{+0.0033}_{-0.0033}  & -18.9789^{+0.0030}_{-0.0030} & 0.5001^{+0.0007}_{-0.0007} & 0.08 & -0.08 & 0.03 & 2.2^{+0.1}_{-0.1} & 70.2^{+0.3}_{-0.3} \\
NGC 6101 & $\ldots$ & 246.450485 & -72.202194 & 1.7300^{+0.0500}_{-0.0500}  & -0.4700^{+0.0500}_{-0.0500} & . $\ldots$ & . $\ldots$ & . $\ldots$ & . $\ldots$ & 15.1^{+0.3}_{-0.4} & 361.4^{+1.7}_{-1.7} \\
NGC 6144 & $\ldots$ & 246.8061 & -26.0301 & -1.7646^{+0.0085}_{-0.0085}  & -2.6371^{+0.0063}_{-0.0063} & 0.0668^{+0.0040}_{-0.0040} & 0.08 & -0.15 & 0.15 & 10.1^{+0.2}_{-0.2} & 188.9^{+1.1}_{-1.1} \\
NGC 6171 & M 107 & 248.135 & -13.057 & -1.9359^{+0.0064}_{-0.0064}  & -5.9487^{+0.0048}_{-0.0048} & 0.1480^{+0.0026}_{-0.0026} & 0.03 & -0.13 & 0.17 & 6.3^{+0.1}_{-0.1} & -33.8^{+0.3}_{-0.3} \\
NGC 6205 & M 13 & 250.4217 & 36.4596 & -3.1762^{+0.0027}_{-0.0027}  & -2.5876^{+0.0030}_{-0.0030} & 0.0801^{+0.0007}_{-0.0007} & 0.19 & -0.04 & 0.04 & 7.0^{+0.2}_{-0.2} & -246.6^{+0.9}_{-0.9} \\
NGC 6218 & M 12 & 251.8101 & -1.951 & -0.1577^{+0.0040}_{-0.0040}  & -6.7683^{+0.0027}_{-0.0027} & 0.1563^{+0.0013}_{-0.0013} & 0.29 & -0.06 & 0.11 & 4.7^{+0.1}_{-0.1} & -42.1^{+0.6}_{-0.6} \\
NGC 6254 & M 10 & 254.2861 & -4.0981 & -4.7031^{+0.0039}_{-0.0039}  & -6.5285^{+0.0027}_{-0.0027} & 0.1511^{+0.0014}_{-0.0014} & 0.21 & -0.07 & 0.1 & 4.3^{+0.1}_{-0.1} & 75.8^{+1.0}_{-1.0} \\
NGC 6304 & $\ldots$ & 258.637 & -29.4816 & -3.9478^{+0.0095}_{-0.0095}  & -1.1248^{+0.0069}_{-0.0069} & 0.1077^{+0.0034}_{-0.0034} & 0.16 & -0.04 & 0.13 & 6.0^{+0.1}_{-0.1} & -107.3^{+3.6}_{-3.6} \\
NGC 6341 & M 92 & 259.2821 & 43.1352 & -4.9367^{+0.0040}_{-0.0040}  & -0.5559^{+0.0040}_{-0.0040} & 0.0564^{+0.0008}_{-0.0008} & 0.11 & -0.04 & -0.0 & 8.1^{+0.2}_{-0.2} & -121.6^{+1.5}_{-1.5} \\
NGC 6352 & $\ldots$ & 261.3739 & -48.427 & -2.1889^{+0.0046}_{-0.0046}  & -4.4209^{+0.0036}_{-0.0036} & 0.1543^{+0.0018}_{-0.0018} & 0.22 & -0.12 & 0.13 & 5.6^{+0.1}_{-0.1} & -120.9^{+3.0}_{-3.0} \\
NGC 6366 & $\ldots$ & 261.9393 & -5.0752 & -0.3835^{+0.0054}_{-0.0054}  & -5.1309^{+0.0044}_{-0.0044} & 0.2292^{+0.0022}_{-0.0022} & 0.27 & -0.08 & 0.14 & 3.6^{+0.1}_{-0.1} & -122.3^{+0.5}_{-0.5} \\
NGC 6362 & $\ldots$ & 262.9772 & -67.0492 & -5.5014^{+0.0028}_{-0.0028}  & -4.7417^{+0.0032}_{-0.0032} & 0.0974^{+0.0011}_{-0.0011} & 0.06 & -0.06 & 0.06 & 7.5^{+0.2}_{-0.2} & -13.1^{+0.6}_{-0.6} \\
NGC 6388 & $\ldots$ & 264.0654 & -44.7423 & -1.3548^{+0.0072}_{-0.0072}  & -2.7144^{+0.0061}_{-0.0061} & 0.0482^{+0.0034}_{-0.0034} & 0.11 & -0.12 & 0.18 & 11.5^{+0.3}_{-0.3} & 81.2^{+1.2}_{-1.2} \\
NGC 6397 & $\ldots$ & 265.1697 & -53.6773 & 3.2908^{+0.0026}_{-0.0026}  & -17.5908^{+0.0025}_{-0.0025} & 0.3781^{+0.0007}_{-0.0007} & 0.1 & -0.05 & 0.07 & 2.2^{+0.1}_{-0.1} & 18.9^{+0.1}_{-0.1} \\
NGC 6441 & $\ldots$ & 267.5540 & -37.066 & -2.5394^{+0.0070}_{-0.0070}  & -5.3010^{+0.0057}_{-0.0057} & 0.0403^{+0.0037}_{-0.0037} & 0.17 & -0.18 & 0.17 & 9.7^{+0.2}_{-0.2} & 18.3^{+1.7}_{-1.7} \\
NGC 6496 & $\ldots$ & 269.7677 & -44.2660 & -3.0290^{+0.0057}_{-0.0057}  & -9.1971^{+0.0050}_{-0.0050} & 0.0803^{+0.0031}_{-0.0031} & 0.07 & -0.15 & 0.17 & 11.6^{+0.3}_{-0.3} & -112.7^{+5.7}_{-5.7} \\
NGC 6535 & $\ldots$ & 270.959 & -0.2953 & -4.2101^{+0.0115}_{-0.0115}  & -2.9461^{+0.0108}_{-0.0108} & 0.1294^{+0.0047}_{-0.0047} & 0.1 & 0.1 & -0.02 & 6.8^{+0.2}_{-0.2} & -215.1^{+0.5}_{-0.5} \\
NGC 6541 & $\ldots$ & 271.9827 & -43.7144 & 0.2762^{+0.0054}_{-0.0054}  & -8.7659^{+0.0048}_{-0.0048} & 0.1139^{+0.0025}_{-0.0025} & -0.03 & -0.04 & 0.08 & 7.4^{+0.2}_{-0.2} & -156.2^{+2.7}_{-2.7} \\
NGC 6584 & $\ldots$ & 274.656646 & -52.215778 & -0.0500^{+0.0100}_{-0.0100}  & -7.2200^{+0.0100}_{-0.0100} & . $\ldots$ & -0.15 & . $\ldots$ & . $\ldots$ & 13.0^{+0.3}_{-0.3} & 222.9^{+15.0}_{-15.0} \\
NGC 6637 & M 69 & 277.8342 & -32.3565 & -5.0669^{+0.0104}_{-0.0104}  & -5.8017^{+0.0094}_{-0.0094} & 0.0746^{+0.0032}_{-0.0032} & 0.23 & -0.01 & -0.05 & 8.2^{+0.2}_{-0.2} & 39.9^{+2.8}_{-2.8} \\
NGC 6656 & M 22 & 279.1048 & -23.9102 & 9.8019^{+0.0036}_{-0.0036}  & -5.5643^{+0.0034}_{-0.0034} & 0.2602^{+0.0009}_{-0.0009} & 0.15 & -0.05 & 0.02 & 3.2^{+0.1}_{-0.1} & -148.9^{+0.4}_{-0.4} \\
NGC 6681 & M 70 & 280.802 & -32.2892 & 1.3853^{+0.0076}_{-0.0076}  & -4.7174^{+0.0065}_{-0.0065} & 0.1096^{+0.0038}_{-0.0038} & 0.26 & -0.24 & 0.1 & 8.7^{+0.2}_{-0.2} & 218.6^{+1.2}_{-1.2} \\
NGC 6723 & $\ldots$ & 284.888123 & -36.632248 & 1.0000^{+0.0100}_{-0.0100}  & -2.4200^{+0.0100}_{-0.0100} & . $\ldots$ & 0.11 & . $\ldots$ & . $\ldots$ & 8.6^{+0.2}_{-0.2} & -94.5^{+3.6}_{-3.6} \\
NGC 6752 & $\ldots$ & 287.7175 & -59.9833 & -3.1908^{+0.0018}_{-0.0018}  & -4.0347^{+0.0020}_{-0.0020} & 0.2310^{+0.0011}_{-0.0011} & 0.19 & -0.29 & 0.03 & 3.9^{+0.1}_{-0.1} & -24.5^{+1.9}_{-1.9} \\
NGC 6779 & M 56 & 289.148 & 30.184 & -2.0092^{+0.0051}_{-0.0051}  & 1.6553^{+0.0056}_{-0.0056} & 0.0702^{+0.0015}_{-0.0015} & 0.03 & -0.05 & -0.03 & 9.9^{+0.2}_{-0.2} & -135.7^{+0.8}_{-0.8} \\
Terzan 7 & $\ldots$ & 289.432983 & -34.657722 & -3.0400^{+0.0600}_{-0.0600}  & -1.7100^{+0.0500}_{-0.0500} & . $\ldots$ & . $\ldots$ & . $\ldots$ & . $\ldots$ & 23.0^{+0.5}_{-0.5} & 166.0^{+4.0}_{-4.0} \\
Arp 2 & $\ldots$ & 292.183807 & -30.355638 & -2.4000^{+0.0400}_{-0.0400}  & -1.5400^{+0.0300}_{-0.0300} & . $\ldots$ & . $\ldots$ & . $\ldots$ & . $\ldots$ & 27.6^{+0.6}_{-0.6} & 115.0^{+10.0}_{-10.0} \\
NGC 6809 & M 55 & 295.0046 & -30.9621 & -3.4017^{+0.0031}_{-0.0031}  & -9.2642^{+0.0028}_{-0.0028} & 0.1707^{+0.0011}_{-0.0011} & 0.18 & -0.03 & 0.0 & 5.3^{+0.1}_{-0.1} & 174.8^{+0.4}_{-0.4} \\
Terzan 8 & $\ldots$ & 295.435028 & -33.999474 & -2.9100^{+0.0800}_{-0.0800}  & -1.6300^{+0.0600}_{-0.0600} & . $\ldots$ & . $\ldots$ & . $\ldots$ & . $\ldots$ & 25.4^{+0.6}_{-0.6} & 130.0^{+8.0}_{-8.0} \\
NGC 6838 & M 71 & 298.4427 & 18.779 & -3.3842^{+0.0027}_{-0.0027}  & -2.6528^{+0.0028}_{-0.0028} & 0.2252^{+0.0010}_{-0.0010} & 0.11 & -0.11 & -0.01 & 3.8^{+0.1}_{-0.1} & -22.9^{+0.2}_{-0.2} \\
NGC 6934 & $\ldots$ & 308.547393 & 7.404472 & -2.6700^{+0.0400}_{-0.0400}  & -4.5200^{+0.0500}_{-0.0500} & . $\ldots$ & . $\ldots$ & . $\ldots$ & . $\ldots$ & 15.2^{+0.3}_{-0.4} & -411.4^{+1.6}_{-1.6} \\
NGC 6981 & M 72 & 313.3662 & -12.5386 & -1.2488^{+0.0089}_{-0.0089}  & -3.3117^{+0.0068}_{-0.0068} & 0.0225^{+0.0063}_{-0.0063} & 0.26 & -0.38 & -0.13 & 16.8^{+0.4}_{-0.4} & -345.1^{+3.7}_{-3.7} \\
NGC 7078 & M 15 & 322.4949 & 12.1661 & -0.6238^{+0.0041}_{-0.0041}  & -3.7960^{+0.0039}_{-0.0039} & 0.0568^{+0.0014}_{-0.0014} & -0.04 & -0.02 & -0.15 & 10.2^{+0.2}_{-0.2} & -107.5^{+0.3}_{-0.3} \\
NGC 7089 & M 2 & 323.3497 & -0.8177 & 3.4911^{+0.0077}_{-0.0077}  & -2.1501^{+0.0071}_{-0.0071} & 0.0591^{+0.0035}_{-0.0035} & -0.04 & -0.14 & -0.14 & 11.4^{+0.3}_{-0.3} & -5.3^{+2.0}_{-2.0} \\
NGC 7099 & M 30 & 325.0888 & -23.1792 & -0.7017^{+0.0063}_{-0.0063}  & -7.2218^{+0.0055}_{-0.0055} & 0.0746^{+0.0040}_{-0.0040} & 0.3 & -0.29 & -0.27 & 7.9^{+0.2}_{-0.2} & -184.2^{+1.0}_{-1.0} \\
Pal 12 & $\ldots$ & 326.661804 & -21.252611 & -3.0600^{+0.0500}_{-0.0500}  & -3.3200^{+0.0500}_{-0.0500} & . $\ldots$ & . $\ldots$ & . $\ldots$ & . $\ldots$ & 18.7^{+0.4}_{-0.4} & 27.8^{+1.5}_{-1.5}
\enddata
\tablecomments{Input globular cluster astrometry, distance,
and radial velocity data.  While most cluster astrometric data
come from \citet{Gaia_2018b}, the data for NGC 6584 and NGC 6723
come from \citet{Baumgardt_2019} while the data for NGC 1261, NGC
4147, NGC 6101, Terzan 7, Arp 2, Terzan 8, NGC 6934, and Pal 12
come from \citet{Sohn_2018}.  Distance and radial velocity data
come from the December 2010 revision of the \citet{Harris_2010}
compilation.\label{tab:input}}
\end{deluxetable*}
\end{longrotatetable}

\clearpage
\begin{longrotatetable}
\begin{deluxetable*}{llDDDDD}
\tabletypesize{\normalsize}
\tablenum{2}
\tablecaption{Age, Metallicity, and Specific Orbital Energy Inferences}
\tablehead{
\colhead{NGC} &
\colhead{Alternate} &
\twocolhead{$\tau$} &
\twocolhead{[M/H]} &
\twocolhead{SOE$_{\text{\texttt{MW14}}}$} &
\twocolhead{SOE$_{\text{\texttt{ScaledMW14}}}$} &
\twocolhead{SOE$_{\text{\texttt{Mc17}}}$} \\
\colhead{Name} &
\colhead{Name} &
\twocolhead{} &
\twocolhead{} &
\twocolhead{[$10^4~\mathrm{km}^2~\mathrm{s}^{-2}$]} &
\twocolhead{[$10^4~\mathrm{km}^2~\mathrm{s}^{-2}$]} &
\twocolhead{[$10^4~\mathrm{km}^2~\mathrm{s}^{-2}$]}}
\decimals
\startdata
NGC 104 & 47 Tuc & 1.02^{+0.07}_{-0.07} & -0.64^{+0.05}_{-0.05} & -7.64^{+0.06}_{-0.05} & -11.82^{+0.05}_{-0.05} & -16.62^{+0.05}_{-0.05} \\
NGC 288 & $\ldots$ & 0.83^{+0.03}_{-0.03} & -0.92^{+0.05}_{-0.05} & -9.04^{+0.08}_{-0.04} & -13.06^{+0.07}_{-0.05} & -16.00^{+0.11}_{-0.07} \\
NGC 362 & $\ldots$ & 0.81^{+0.04}_{-0.04} & -0.87^{+0.05}_{-0.05} & -8.66^{+0.16}_{-0.18} & -12.84^{+0.17}_{-0.17} & -16.20^{+0.17}_{-0.20} \\
NGC 1261 & $\ldots$ & 0.80^{+0.04}_{-0.04} & -0.86^{+0.05}_{-0.05} & -8.20^{+0.07}_{-0.06} & -12.04^{+0.06}_{-0.08} & -13.43^{+0.13}_{-0.09} \\
NGC 1851 & $\ldots$ & 0.78^{+0.04}_{-0.04} & -0.81^{+0.05}_{-0.05} & -7.57^{+0.15}_{-0.10} & -11.34^{+0.14}_{-0.09} & -13.47^{+0.14}_{-0.09} \\
NGC 2298 & $\ldots$ & 0.99^{+0.05}_{-0.05} & -1.49^{+0.05}_{-0.05} & -8.04^{+0.19}_{-0.10} & -11.85^{+0.17}_{-0.13} & -14.15^{+0.22}_{-0.15} \\
NGC 2808 & $\ldots$ & 0.85^{+0.03}_{-0.03} & -0.89^{+0.05}_{-0.05} & -8.12^{+0.08}_{-0.07} & -12.24^{+0.13}_{-0.08} & -15.33^{+0.11}_{-0.10} \\
NGC 3201 & $\ldots$ & 0.80^{+0.03}_{-0.03} & -1.02^{+0.05}_{-0.05} & -3.13^{+0.19}_{-0.25} & -7.21^{+0.19}_{-0.25} & -11.01^{+0.23}_{-0.26} \\
NGC 4147 & $\ldots$ & 0.89^{+0.04}_{-0.04} & -1.28^{+0.05}_{-0.05} & -7.34^{+0.04}_{-0.04} & -11.08^{+0.06}_{-0.04} & -12.33^{+0.11}_{-0.09} \\
NGC 4590 & M 68 & 0.90^{+0.04}_{-0.04} & -1.78^{+0.05}_{-0.05} & -3.94^{+0.56}_{-0.59} & -8.17^{+0.71}_{-0.56} & -11.41^{+0.51}_{-0.55} \\
NGC 4833 & $\ldots$ & 0.98^{+0.05}_{-0.05} & -1.49^{+0.05}_{-0.05} & -9.37^{+0.10}_{-0.09} & -13.58^{+0.08}_{-0.08} & -18.66^{+0.12}_{-0.11} \\
NGC 5024 & M 53 & 0.99^{+0.05}_{-0.05} & -1.64^{+0.05}_{-0.05} & -7.44^{+0.88}_{-0.53} & -11.04^{+0.97}_{-0.70} & -12.26^{+0.90}_{-0.53} \\
NGC 5053 & $\ldots$ & 0.96^{+0.04}_{-0.04} & -1.76^{+0.05}_{-0.05} & -7.75^{+0.73}_{-0.60} & -11.68^{+0.92}_{-0.60} & -12.97^{+0.75}_{-0.60} \\
NGC 5139 &  $\omega$ Cen & 0.90^{+0.05}_{-0.05} & -1.13^{+0.05}_{-0.05} & -9.29^{+0.11}_{-0.06} & -13.51^{+0.07}_{-0.08} & -18.60^{+0.12}_{-0.07} \\
NGC 5272 & M 3 & 0.89^{+0.04}_{-0.04} & -1.12^{+0.05}_{-0.05} & -7.55^{+0.13}_{-0.12} & -11.61^{+0.13}_{-0.12} & -14.06^{+0.13}_{-0.12} \\
NGC 5286 & $\ldots$ & 0.98^{+0.04}_{-0.04} & -1.19^{+0.05}_{-0.05} & -8.06^{+0.07}_{-0.08} & -12.48^{+0.11}_{-0.07} & -15.88^{+0.16}_{-0.13} \\
NGC 5466 & $\ldots$ & 1.06^{+0.05}_{-0.05} & -1.98^{+0.05}_{-0.05} & -3.35^{+0.65}_{-0.52} & -7.27^{+0.44}_{-0.53} & -9.00^{+0.73}_{-0.46} \\
NGC 5904 & M 5 & 0.83^{+0.03}_{-0.03} & -0.90^{+0.05}_{-0.05} & -4.44^{+0.59}_{-0.39} & -8.76^{+0.50}_{-0.41} & -12.72^{+0.68}_{-0.43} \\
NGC 5927 & $\ldots$ & 0.99^{+0.07}_{-0.07} & -0.50^{+0.05}_{-0.05} & -7.18^{+0.09}_{-0.10} & -11.62^{+0.10}_{-0.09} & -18.37^{+0.10}_{-0.07} \\
NGC 5986 & $\ldots$ & 0.95^{+0.04}_{-0.04} & -1.13^{+0.05}_{-0.05} & -10.81^{+0.19}_{-0.16} & -15.47^{+0.21}_{-0.12} & -20.28^{+0.29}_{-0.22} \\
NGC 6093 & M 80 & 0.98^{+0.04}_{-0.04} & -1.25^{+0.05}_{-0.05} & -10.84^{+0.07}_{-0.05} & -15.42^{+0.08}_{-0.07} & -21.78^{+0.17}_{-0.12} \\
NGC 6121 & M 4 & 0.98^{+0.05}_{-0.05} & -0.83^{+0.05}_{-0.05} & -9.72^{+0.02}_{-0.01} & -13.93^{+0.02}_{-0.01} & -19.69^{+0.05}_{-0.04} \\
NGC 6101 & $\ldots$ & 0.98^{+0.04}_{-0.04} & -1.54^{+0.05}_{-0.05} & -5.04^{+0.13}_{-0.11} & -9.60^{+0.13}_{-0.11} & -10.27^{+0.22}_{-0.26} \\
NGC 6144 & $\ldots$ & 1.08^{+0.05}_{-0.05} & -1.52^{+0.05}_{-0.05} & -9.58^{+0.18}_{-0.18} & -14.31^{+0.15}_{-0.13} & -19.82^{+0.18}_{-0.20} \\
NGC 6171 & M 107 & 1.09^{+0.06}_{-0.06} & -0.81^{+0.05}_{-0.05} & -10.00^{+0.09}_{-0.09} & -14.44^{+0.08}_{-0.07} & -21.27^{+0.11}_{-0.08} \\
NGC 6205 & M 13 & 0.91^{+0.04}_{-0.04} & -1.11^{+0.05}_{-0.05} & -9.32^{+0.10}_{-0.07} & -13.51^{+0.10}_{-0.05} & -17.34^{+0.11}_{-0.08} \\
NGC 6218 & M 12 & 0.99^{+0.03}_{-0.03} & -0.92^{+0.05}_{-0.05} & -9.01^{+0.15}_{-0.18} & -13.32^{+0.18}_{-0.19} & -19.47^{+0.17}_{-0.20} \\
NGC 6254 & M 10 & 0.89^{+0.04}_{-0.04} & -1.03^{+0.05}_{-0.05} & -8.82^{+0.10}_{-0.08} & -13.14^{+0.11}_{-0.07} & -19.37^{+0.12}_{-0.08} \\
NGC 6304 & $\ldots$ & 1.06^{+0.08}_{-0.08} & -0.52^{+0.05}_{-0.05} & -8.08^{+0.23}_{-0.21} & -12.52^{+0.22}_{-0.22} & -22.01^{+0.27}_{-0.34} \\
NGC 6341 & M 92 & 1.03^{+0.04}_{-0.04} & -1.94^{+0.05}_{-0.05} & -9.00^{+0.10}_{-0.11} & -13.16^{+0.13}_{-0.10} & -16.63^{+0.09}_{-0.11} \\
NGC 6352 & $\ldots$ & 0.99^{+0.07}_{-0.07} & -0.56^{+0.05}_{-0.05} & -7.69^{+0.17}_{-0.15} & -12.06^{+0.13}_{-0.19} & -20.12^{+0.18}_{-0.16} \\
NGC 6366 & $\ldots$ & 1.04^{+0.13}_{-0.13} & -0.59^{+0.05}_{-0.05} & -8.53^{+0.19}_{-0.10} & -12.78^{+0.15}_{-0.11} & -19.31^{+0.20}_{-0.10} \\
NGC 6362 & $\ldots$ & 1.06^{+0.05}_{-0.05} & -0.85^{+0.05}_{-0.05} & -8.97^{+0.08}_{-0.07} & -13.36^{+0.07}_{-0.09} & -19.02^{+0.09}_{-0.11} \\
NGC 6388 & $\ldots$ & 0.94^{+0.08}_{-0.08} & -0.63^{+0.05}_{-0.05} & -11.34^{+0.21}_{-0.16} & -16.13^{+0.27}_{-0.18} & -20.99^{+0.26}_{-0.29} \\
NGC 6397 & $\ldots$ & 0.99^{+0.04}_{-0.04} & -1.54^{+0.05}_{-0.05} & -8.24^{+0.04}_{-0.06} & -12.44^{+0.05}_{-0.06} & -18.30^{+0.04}_{-0.06} \\
NGC 6441 & $\ldots$ & 0.88^{+0.07}_{-0.07} & -0.46^{+0.05}_{-0.05} & -11.26^{+0.09}_{-0.08} & -16.03^{+0.11}_{-0.08} & -24.53^{+0.40}_{-0.33} \\
NGC 6496 & $\ldots$ & 0.97^{+0.07}_{-0.07} & -0.56^{+0.05}_{-0.05} & -5.86^{+0.60}_{-0.60} & -10.94^{+0.69}_{-0.43} & -15.40^{+0.77}_{-0.79} \\
NGC 6535 & $\ldots$ & 0.82^{+0.09}_{-0.09} & -1.29^{+0.05}_{-0.05} & -9.38^{+0.09}_{-0.09} & -13.84^{+0.12}_{-0.11} & -20.81^{+0.08}_{-0.08} \\
NGC 6541 & $\ldots$ & 1.01^{+0.04}_{-0.04} & -1.31^{+0.05}_{-0.05} & -7.52^{+0.11}_{-0.13} & -12.05^{+0.15}_{-0.14} & -20.74^{+0.12}_{-0.13} \\
NGC 6584 & $\ldots$ & 0.88^{+0.03}_{-0.03} & -1.10^{+0.05}_{-0.05} & -4.81^{+0.78}_{-0.75} & -9.54^{+0.64}_{-0.73} & -11.81^{+0.92}_{-0.91} \\
NGC 6637 & M 69 & 1.02^{+0.07}_{-0.07} & -0.64^{+0.05}_{-0.05} & -10.39^{+0.17}_{-0.17} & -14.94^{+0.17}_{-0.16} & -24.16^{+0.21}_{-0.17} \\
NGC 6656 & M 22 & 0.99^{+0.05}_{-0.05} & -1.27^{+0.05}_{-0.05} & -5.39^{+0.15}_{-0.16} & -9.63^{+0.19}_{-0.19} & -16.48^{+0.15}_{-0.17} \\
NGC 6681 & M 70 & 1.00^{+0.04}_{-0.04} & -1.13^{+0.05}_{-0.05} & -7.31^{+0.42}_{-0.35} & -11.96^{+0.45}_{-0.34} & -19.70^{+0.45}_{-0.34} \\
NGC 6723 & $\ldots$ & 1.02^{+0.05}_{-0.05} & -0.82^{+0.05}_{-0.05} & -9.40^{+0.43}_{-0.27} & -14.01^{+0.42}_{-0.29} & -21.21^{+0.49}_{-0.35} \\
NGC 6752 & $\ldots$ & 0.92^{+0.04}_{-0.04} & -1.02^{+0.05}_{-0.05} & -8.18^{+0.13}_{-0.09} & -12.41^{+0.08}_{-0.10} & -18.53^{+0.13}_{-0.10} \\
NGC 6779 & M 56 & 1.07^{+0.05}_{-0.05} & -1.50^{+0.05}_{-0.05} & -8.00^{+0.17}_{-0.14} & -12.27^{+0.17}_{-0.16} & -15.73^{+0.18}_{-0.17} \\
Terzan 7 & $\ldots$ & 0.57^{+0.04}_{-0.04} & -0.42^{+0.05}_{-0.05} & -9.39^{+0.21}_{-0.19} & -14.83^{+0.22}_{-0.18} & -9.46^{+0.34}_{-0.35} \\
Arp 2 & $\ldots$ & 0.85^{+0.06}_{-0.06} & -1.23^{+0.05}_{-0.05} & -9.59^{+0.43}_{-0.33} & -14.87^{+0.35}_{-0.33} & -9.10^{+0.44}_{-0.36} \\
NGC 6809 & M 55 & 0.96^{+0.04}_{-0.04} & -1.32^{+0.05}_{-0.05} & -8.19^{+0.05}_{-0.03} & -12.56^{+0.05}_{-0.04} & -18.88^{+0.05}_{-0.04} \\
Terzan 8 & $\ldots$ & 0.95^{+0.04}_{-0.04} & -1.58^{+0.05}_{-0.05} & -8.12^{+0.40}_{-0.37} & -13.22^{+0.34}_{-0.39} & -8.35^{+0.56}_{-0.35} \\
NGC 6838 & M 71 & 1.07^{+0.08}_{-0.08} & -0.59^{+0.05}_{-0.05} & -7.69^{+0.04}_{-0.05} & -11.88^{+0.05}_{-0.05} & -17.25^{+0.04}_{-0.04} \\
NGC 6934 & $\ldots$ & 0.87^{+0.04}_{-0.04} & -1.10^{+0.05}_{-0.05} & -4.97^{+0.20}_{-0.24} & -9.37^{+0.22}_{-0.23} & -10.43^{+0.33}_{-0.37} \\
NGC 6981 & M 72 & 0.85^{+0.02}_{-0.02} & -0.99^{+0.05}_{-0.05} & -7.58^{+0.67}_{-0.45} & -12.04^{+0.71}_{-0.45} & -12.66^{+0.79}_{-0.51} \\
NGC 7078 & M 15 & 1.01^{+0.04}_{-0.04} & -1.80^{+0.05}_{-0.05} & -9.24^{+0.09}_{-0.06} & -13.43^{+0.09}_{-0.06} & -16.15^{+0.13}_{-0.13} \\
NGC 7089 & M 2 & 0.92^{+0.04}_{-0.04} & -1.09^{+0.05}_{-0.05} & -7.14^{+0.42}_{-0.39} & -11.54^{+0.63}_{-0.42} & -13.97^{+0.38}_{-0.39} \\
NGC 7099 & M 30 & 1.01^{+0.05}_{-0.05} & -1.70^{+0.05}_{-0.05} & -9.24^{+0.21}_{-0.12} & -13.56^{+0.23}_{-0.13} & -17.53^{+0.29}_{-0.17} \\
Pal 12 & $\ldots$ & 0.69^{+0.09}_{-0.09} & -0.69^{+0.05}_{-0.05} & -5.12^{+0.24}_{-0.26} & -9.48^{+0.27}_{-0.24} & -9.18^{+0.34}_{-0.35}
\enddata
\tablecomments{Table \ref{tab:inference} is ordered by R.A.  The first
column is normalized age $\tau = \text{age}/12.8\text{~Gyr}$ from
\citet{Marin_2009} calculated using the Dartmouth Stellar Evolution
Program \citep{Chaboyer_2001,Bjork_2006,Dotter_2007} assuming the
metallicities presented in \citet{Rutledge_1997b,Rutledge_1997a} on the
\citet{Carretta_1997} scale.  The second column metallicity [M/H] is from
\citet{Rutledge_1997b,Rutledge_1997a} on the \citet{Carretta_1997} scale.
The third, fourth, and fifth columns are globular cluster specific orbital
energies calculated using the data in Table \ref{tab:input} assuming the
default, \texttt{MWPotential2014}, the scaled \texttt{MWPotential2014}
\citep{Bovy_2015}, and the \texttt{McMillan17} \citep{McMillan_2017}
potentials respectively.\label{tab:inference}}
\end{deluxetable*}
\end{longrotatetable}

\end{document}